\documentclass[twocolumn]{aastex631} 

\usepackage{natbib}

\usepackage{gensymb} 
\usepackage{amsmath} 
\usepackage{graphicx}

\newcommand{\HI}{\ensuremath{\mbox{\ion{H}{1}}}}
\newcommand{\OVI}{\ensuremath{\mbox{\ion{O}{6}}}}
\newcommand{\SiII}{\ensuremath{\mbox{\ion{Si}{2}}}}

\newcommand{\SII}{\ensuremath{\mbox{\ion{S}{2}}}}
\newcommand{\CIV}{\ensuremath{\mbox{\ion{C}{4}}}}
\newcommand{\SiIV}{\ensuremath{\mbox{\ion{Si}{4}}}}
\newcommand{\NV}{\ensuremath{\mbox{\ion{N}{5}}}}

\newcommand{\halpha}{H$\alpha$\relax}

\newcommand{\Nav}{\ensuremath{N_a(v)}}

\newcommand{\e}[1]{10^{#1}}
\newcommand{\kms}{\ensuremath{{\rm km\,s}^{-1}}}
\newcommand{\cIVsiIV}{$N_{\rm C\,IV}/N_{\rm Si\,IV}$}
\newcommand{\cIVsiIVav}{$\langle N_{\rm C\,IV}/N_{\rm Si\,IV} \rangle$}
\newcommand{\siIVsII}{$N_{\rm Si\,IV}/N_{\rm S\,II}$}
\newcommand{\siIVsIIav}{$\langle\log N_{\rm Si\,IV}/N_{\rm S\,II} \rangle$}

\newcommand{\vlsr}{\ensuremath{v_{\rm LSR}\relax}}

\newcommand{\z}{$|z_{\rm P}|$}
\defcitealias{Savage2009}{SW09}
\defcitealias{Zheng2019}{Z19}
\defcitealias{Lehner2020}{M31}
\defcitealias{Dickey1990}{DL90}
\defcitealias{Haffner1999}{H99}
\defcitealias{Lehner2011}{L11}

\shortauthors{Tuli et al.}
\shorttitle{The Perseus Project}

\begin{document}

\title{The Warm-Hot Disk-Halo Interface Below the Perseus Spiral Arm \footnote{Based on observations made with the NASA/ESA Hubble Space Telescope, obtained from the data archive at the Space Telescope Science Institute. STScI is operated by the Association of Universities for Research in Astronomy, Inc. under NASA contract NAS 5-26555.}}

\author[0000-0001-6398-5466]{Ananya Goon Tuli}
\affiliation{Department of Physics and Astronomy, University of Notre Dame, Notre Dame, IN 46556}

\author[0000-0001-9158-0829]{Nicolas Lehner}
\affiliation{Department of Physics and Astronomy, University of Notre Dame, Notre Dame, IN 46556}

\author[0000-0002-2591-3792]{J. Christopher Howk}
\affiliation{Department of Physics and Astronomy, University of Notre Dame, Notre Dame, IN 46556}

\author[0000-0002-1218-640X]{Todd M. Tripp}
\affiliation{Department of Astronomy, University of Massachusetts, Amherst, MA 01003-9305}

\author[0000-0003-0724-4115]{Andrew J. Fox}
\affil{AURA for ESA, Space Telescope Science Institute, 3700 San Martin Drive, Baltimore, MD 21218}

\author[0000-0003-4237-3553]{      Frances H. Cashman}
\affil{Department of Physics, Presbyterian College, 503 South Broad Street, Clinton, SC 29325}



\begin{abstract}

The Milky Way's disk-halo interface mediates energy and mass exchange between the interstellar thin disk and the halo. In the first detailed study of the Perseus arm's disk-halo interface, we combine HST/STIS and COS absorption spectra toward 6 stars and 23 AGNs projected behind a narrow section ($95\degree<l<145\degree$, $-46\degree<b<0\degree$), providing a unique dataset that bridges the disk and its extended vertical structure in these directions. We measure \SII, \SiIV, and \CIV\ absorption, along with \HI\ 21 cm emission, at heights $-70$ pc to $-3.3$ kpc from the mid-plane. The arm's southern vertical structure shows distinct height-dependent behaviors: \HI\ and \SII\ column densities sharply decline with height up to 1.5 kpc, then continue declining at a much shallower rate at greater heights. In contrast, high ion (Si IV and C IV) column densities remain relatively constant throughout the entire height range. In the disk-halo interface, where warm neutral medium dominates, \SII\ and the high ions show similar kinematics, and we find a remarkably uniform \CIV/\SiIV\ ratio (\cIVsiIVav$\,=2.5 \pm 0.5$) within $-0.9$ to $-3.25$ kpc. Both the kinematic correspondence and high-ion ratio are consistent with the high ions probing turbulent mixing layers at the interfaces between warm/cool and hot gas phases. The AGN sightlines reveal minimal circumgalactic medium (CGM) contribution in the low-velocity gas at $|v_{\rm LSR}|< 100$ \kms. The extraplanar absorbing gas may trace material ejected from previous Galactic fountain activity.

\end{abstract}



\section{Introduction}
\label{sec:intro}
The Milky Way's spiral arms are loci of active star formation \citep{McKee2007,Elmegreen2011}. Within the spiral arms, spatially- and temporally-correlated feedback from stars (e.g., strong stellar winds and supernova explosions) evolve into super-bubbles that expand through the interstellar medium (ISM). Eventually, these super-bubbles break out as Galactic fountains \citep{Shapiro1976, Bregman1980} and chimneys \citep{Norman1989} when the hot ($T>10^6$ K) gas escapes into the extraplanar regions above/below the spiral arms. This activity plays a crucial role in galaxy evolution: it is now widely recognized that the competition between \textit{feedback} (driven by outflows and escaping ionizing light) vs. \textit{accretion} (inflows of relatively pristine gas as well as recycled fountain material) regulates star formation \citep{Thompson2024} and enables galaxies to continue to evolve over their lifetimes. The extraplanar disk-halo interface region, roughly within 5 kpc of the Galactic plane \citep[e.g.,][]{Bowen2008,Savage2009}, is the linchpin of this ``baryon cycle'': this is where outflowing matter, energy, and momentum are coupled with the broader surroundings of a galaxy, and conversely, this is the region that inflowing material must navigate to survive its descent into the disk, where it can eventually form new stars. The disk-halo interface thus connects the Milky Way's interstellar thin disk \citep[scale height $\sim 200$--$300$ pc,][]{Ojha2001} to the distant halo gas known as the circumgalactic medium (CGM) (\citealt{Savage2003,Putman2009, Wakker2012}, Cashman et al. 2025).

The disk-halo interface is supported by the energy and matter of stellar feedback from the spiral arms \citep{Heckman2002, Kim2018}, which impacts the disk-halo interface's large-scale density distribution and extension of various gas phases above the Galactic plane. The bulk transportation of matter from the spiral arms into the disk-halo interface through galactic fountains and chimneys sets the vertical extension of the arms as well as determines the ionization conditions in the extraplanar region. Moreover, the disk-halo interface facilitates the mixing of enriched feedback material with the falling pristine halo material \citep{Putman2009}, which determine the chemical composition of the disk-halo interface gas. Therefore, the disk-halo interface's gas distribution, extension, ionization condition, and chemical composition are important aspects of the Galactic structure and evolution.

Historically, the spatial structure of the nearby disk-halo interface gas has been characterized as a flat slab with an exponential density distribution \citep{Lockman1986, Dickey1990, Savage1985, Sembach1992, Savage1997, Haffner1999, Savage2003, Bowen2008, Savage2009, Wakker2012}. Using this simplified concept, the scale heights of the different disk-halo interface gas phases has been estimated, e.g., $h_z= 0.5$ kpc for the warm neutral medium (WNM) ($\leq10^4$ K) \citep{Lockman1986, Dickey1990}, 1 kpc for the warm ionized medium (WIM) ($\sim 10^{4}$--$10^{4.2}$ K) \citep{Haffner1999, Krishnarao2017}, and 3--4 kpc for the transition temperature phase between $\sim 10^{4.2}$--$10^{5.5}$ K \citep{Savage1985, Sembach1992, Savage1997, Savage2003, Bowen2008, Savage2009, Wakker2012}.

Characterizing the spatial structure of the transition temperature phase in the disk-halo interface is particularly important since the hot feedback material cools through this phase. \SiIV, \CIV, \NV, and \OVI\ ions have some of the strongest transition lines in UV and provide significant cooling rates for the hot material, thus regulating the disk-halo interface's physical conditions. The transition temperature ions' column density distributions observed using halo star sightlines agree with the flat slab model fits \citep{Bowen2008, Savage2009}. However, column densities observed along a large sample of AGN sightlines significantly deviate from the flat-slab model fits \citep{Zheng2019, Qu2019, Qu2020, Qu2022}. Additionally, stellar feedback, e.g., Galactic fountains and chimneys, may substantially alter the gas distribution and ionization conditions above/below the spiral arms---a phenomenon yet unexplored. To resolve the observational tensions between the halo star and AGN sightlines as well as to evaluate stellar feedback's impact on the disk-halo interface gas, we design an experiment investigating the extraplanar \HI, \SII, \SiIV\ and \CIV\ density distribution beneath the Perseus arm using \HI\ 21 cm emission and UV spectra of OB stars and AGNs. The Perseus arm provides unique advantages: its proximity enables precise stellar distance measurements via \textit{Gaia} parallaxes \citep{Gaia}, and its kinematics facilitate unambiguous gas localization. Our methodology (Section \ref{sec:experiment}) directly probes the vertical density distribution and ionization physics of extraplanar Perseus gas at the disk-halo interface, potentially illuminating galactic fountain dynamics.

This paper is organized as follows: Section \ref{sec:experiment}
and Section \ref{sec:sample} describe our experiment and the target selection process. In section \ref{sec:instrument}, we discuss the data products and the analysis process. In Section \ref{sec:Results}, we present our results and discuss their implications in Section \ref{sec:Discussion}. Finally, we summarize our main findings in Section \ref{sec:Summary}.

\begin{deluxetable*}{lcCCcCCCccC}
\tabcolsep=3pt
\tablenum{1} \tabletypesize{\footnotesize}
\tablecaption{Sample Summary \label{tab:sightline}}
\tablehead{\colhead{Target} & \colhead{ID} & \colhead{$l$} & \colhead{$b$} & \colhead{Type} & \colhead{$D_{\rm star}^{\rm a}$} & \colhead{$D_{\rm P}^{\rm b}$} & \colhead{$|z_{\rm P}|^{\rm c}$} & \colhead{Instruments} & \colhead{$S/N$$^{\rm d}$} \\ \colhead{ } & \colhead{ } & \colhead{($\degree$)} & \colhead{($\degree$)} & \colhead{ } & \colhead{(kpc)} & \colhead{(kpc)} & \colhead{(kpc)} & \colhead{ } & \colhead{\CIV/\SiIV}}
\startdata
BD+60~73 & S1 & 121.22 & $-1.46$ & B1Ib & 4.00\; $(^{+0.20}_{-0.18})$ & 2.60 & 0.07 &  STIS E140M & 10/9 \\
HD 14434 & S2 & 135.08 & $-3.82$ &  O5.5Vnn(f)p & 2.52\; $(^{+0.15}_{-0.14}$) & 2.24 & 0.15 &  STIS E140M & 62/49 \\
HD 13745 & S3 & 134.58 & $-4.96$ &  O9.7II(n) & 2.60\; $(^{+0.19}_{-0.16})$ & 2.25 & 0.21 &  STIS E140M & 57/62 \\
HD 210809 & S4 & 99.85 & $-3.13$ &  O9Iab & 4.33\; $(^{+0.73}_{-0.54})$ & 3.69 & 0.21 &  STIS E140H/E140M & 84/21  \\
HD 232522 & S5 & 130.70 & $-6.71$ &  B1II & 4.12\; $(^{+0.74}_{-0.55})$ & 2.33 & 0.27 &  STIS E140H & \nodata/274 \\
3C66A & A1 & 140.14 & $-16.77$ &  AGN & \nodata & 2.15 & 0.65 &  COS G130M/G160M & 7/11 \\
Zw535.012 & A2 & 120.17 & $-17.13$ &  AGN & \nodata & 2.63 & 0.81 &  COS G130M/G160M & 5/8 \\
FBS0150+396 & A3 & 135.58 & $-21.42$ &  AGN & \nodata & 2.23 & 0.87 &  COS G130M/G160M & 7/5 \\
IRASF00040+4325 & A4 & 114.42 & $-18.42$ &  AGN & \nodata & 2.86 & 0.92 &  COS G130M/G160M & 8/15 \\
UVQSJ001903.85+423809.0 & A5 & 116.61 & $-19.84$ &  AGN & \nodata & 2.77 & 0.99 &  COS G130M/G160M & 4/5 \\
PGC2304 & A6 & 120.34 & $-21.33$ &  AGN & \nodata & 2.63 & 1.03 &  COS G130M/G160M & 4/4 & \\
RXSJ0118.8+3836 & A7 & 128.78 & $-23.95$ &  AGN & \nodata & 2.38 & 1.06 &  COS G130M/G160M & 5/5 \\
IVZw29 & A8 & 121.04 & $-22.51$ &  AGN & \nodata & 2.60 & 1.08 &  COS G130M/G160M & 11/12  \\
RXJ0048.3+3941 & A9 & 122.28 & $-23.18$ &  AGN & \nodata & 2.56 & 1.11 &  COS G130M/G160M & 16/6 \\
RXJ0049.8+3931 & A10 & 122.60 & $-23.35$ &  AGN & \nodata & 2.55 & 1.11 &  COS G130M/G160M & 8/60 \\
LAMOSTJ003432.52+391836.1 & A11 & 119.37 & $-23.44$ &  AGN & \nodata & 2.66 & 1.15 &  COS G130M/G160M & 4/6 \\
RXSJ0155.6+3115 & A12 & 138.70 & $-29.64$ &  AGN & \nodata & 2.17 & 1.23 &  COS G130M/G160M & 4/10 \\
RXJ0043.6+3725 & A13 & 121.23 & $-25.42$ &  AGN & \nodata & 2.60 & 1.24 &  COS G130M/G160M & 7/8 \\
3C48.0 & A14 & 133.96 & $-28.72$ &  AGN & \nodata & 2.26 & 1.24 &  COS G130M/G160M & 5/7 \\
RXJ0050.8+3536 & A15 & 122.80 & $-27.26$ &  AGN & \nodata & 2.55 & 1.31 &  COS G130M/G160M & 7/7 \\
RXJ0028.1+3103 & A16 & 117.09 & $-31.54$ &  AGN & \nodata & 2.75 & 1.69 &  COS G130M/G160M & 7/15 \\
PG0052+251 & A17 & 123.91 & $-37.44$ &  AGN & \nodata & 2.51 & 1.92 &  COS G130M/G160M & 8/18 \\
PG0122+214 & S6 & 133.37 & $-40.57$ &  B2-3V & 8.96\; (^{+6.06}_{-2.58}) & 2.27 & 1.94 &  STIS E140M & 10/15 \\
4C25.01 & A18 & 114.07 & $ -36.28$ &  AGN & \nodata & 2.87 & 2.11 &  COS G130M/G160M & 11/8 \\
RXJ0053.7+2232 & A19 & 123.64 & $-40.33$ &  AGN & \nodata & 2.52 & 2.14 &  COS G130M/G160M & 4/8 \\
RBS2055 & A20 & 106.67 & $-34.66$ &  AGN & \nodata & 3.24 & 2.24 &  COS G130M/G160M & 8/9 \\
MRK1148 & A21 & 123.09 & $-45.44$ &  AGN & \nodata & 2.54 & 2.58 &  COS G130M/G160M & 7/9 \\
MRK335 & A22 & 108.76 & $-41.42$ &  AGN & \nodata & 3.13 & 2.76 &  COS G130M/G160M & 14/15 \\
PG0003+158 & A23 & 107.32 & $-45.33$ &  AGN & \nodata & 3.21 & 3.25 &  COS G130M/G160M & 10/11 \\
\enddata
\tablecomments{(a): Distances to the stars with $68\%$ confidence interval using zero-point bias corrected Gaia parallaxes \citep{Gaia}. (b): Distance to the Perseus arm in the direction of the target using the spiral arm equation and the parameters for the Perseus Arm in Table 2 of \citet{Reid2014}. (c): mid-arm height \z\ below the Perseus arm. (d): Signal-to-noise ratio per resolution derived in the stellar continuum nearer the \CIV\ and \SiIV\ lines, respectively. Stellar spectral type references: \citealt{Morgan1955}; \citealt{Walborn1972}; \citealt{Walborn1973}; \citealt{Humphreys1978} \citealt{Ramspeck2001};\citealt{Mariz2004} }
\end{deluxetable*}

\section{Our Experiment}
\label{sec:experiment}

\begin{figure}[h]
\fig{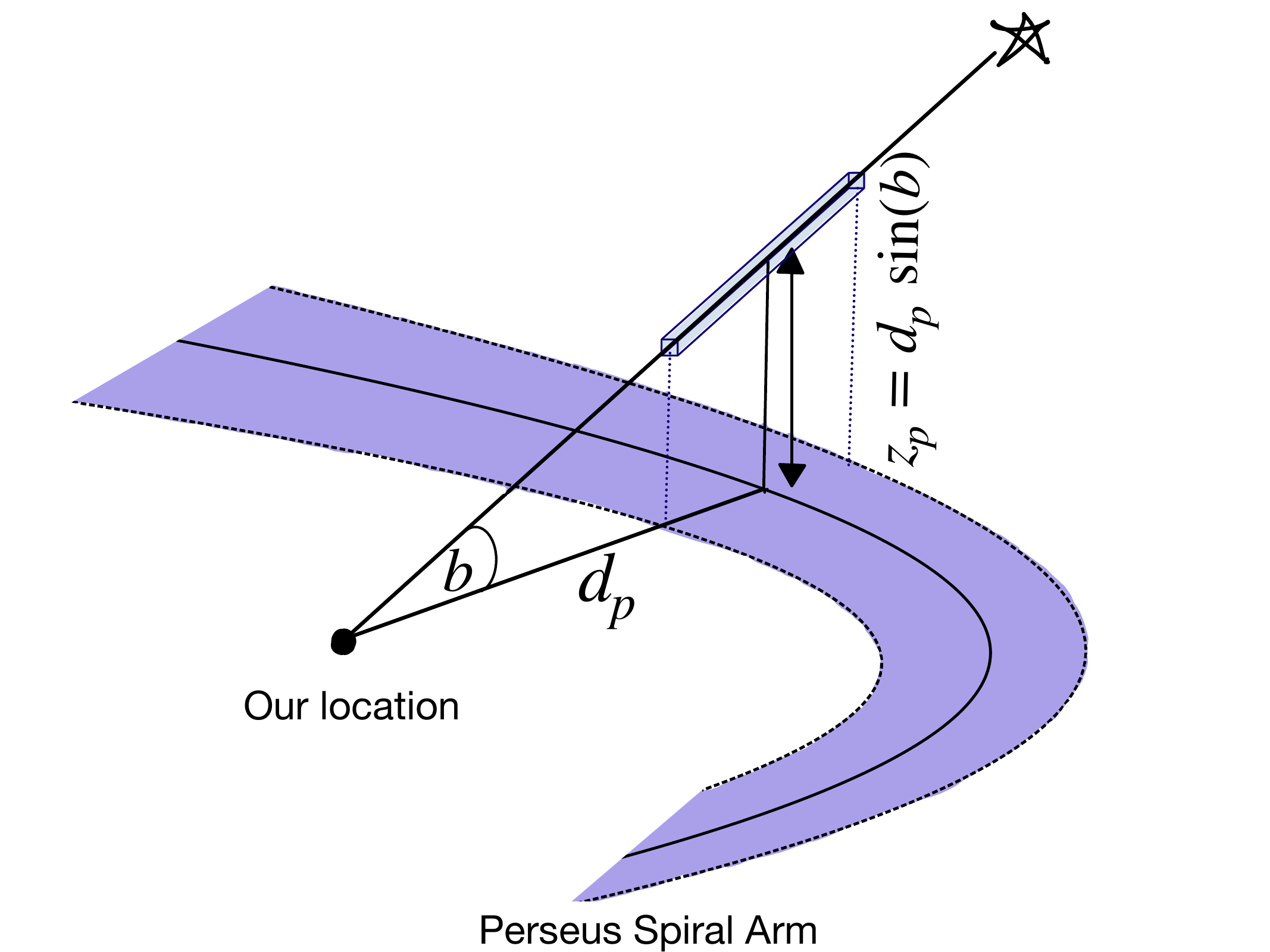}{0.95\columnwidth}{}
  \caption{A schematic of our experiment. A sightline toward a target projected behind the Perseus arm passes above/below the Perseus arm at a specific mid-arm height \z. Based on kinematics, we select the absorption associated with the extraplanar Perseus gas (the blue parallelogram) above/below the arm (see Section~\ref{subsec:velocity} and Table \ref{tab:vel-range}) and measure its column density (see Section~\ref{subsection:column} and Table \ref{tab:column}). Therefore, we can estimate the column density vs. \z\ distribution using a sample of multiple sightlines (see Section~\ref{subsec:column_dist} and Figure~\ref{fig:colden-zheight}). (Art-credit: Saloni Deepak)
  \label{fig:experiment}}
\end{figure}

Figure~\ref{fig:experiment} illustrates our experimental design for probing the vertical structure of the Perseus arm. We utilize absorption spectroscopy along sightlines toward background sources projected behind the Perseus arm. Each sightline intersects the arm's extraplanar gas (depicted by the parallelogram), spanning a range of heights. The mid-arm height of the extraplanar gas (above/below the Galactic plane) along a sightline is $|z_{\rm{P}}| = d_{\rm P} \sin b $, where $d_{\rm P}$ is the mid-arm distance to the Perseus arm in the sightline's direction, and $b$ is the latitude of the sightline. This extraplanar gas imprints characteristic absorption features in the spectra of background sources. Knowing the gas kinematics (see Section \ref{sec:sample}), we can isolate Perseus absorption in these spectra to derive the associated column densities. Thus, using a set of background sources projected behind the Perseus arm and passing above/below it at a range of \z\ values, we can probe the spatial structure and physical conditions in the arm's vertical extension. We conduct this experiment over a narrow section of the Perseus arm where its spiral structure and gas kinematics are well understood (see Section~\ref{sec:sample}), ensuring appropriate identification of the Perseus absorption toward each sightline and thus reliable characterization of the arm's vertical structure.

Our experiment differs from previous multi-sightline surveys \citep[e.g.,][]{Diplas1994, Savage1997, Savage2003, Bowen2008, Savage2009, Wakker2012} in two fundamental ways. The first distinction lies in our column density selection method. While previous works measured the integrated total column density ($N$) along the entire sightline, we use kinematic information to isolate and measure only the column density of the extraplanar Perseus arm gas. The second key difference is in our vertical distance metric. Earlier studies assessed the disk-halo interface gas distribution using sightlines through out the Galaxy, and measuring the total vertical column density ($N\sin{b}$) integrated up to the maximum height, $z$, of each sightline. Our approach, however, examines the column density associated with the Perseus arm at a specific \z, the Perseus mid-arm height---that is we are measuring $N(z=z_{\rm P})$ instead of $N(<z) = \int N(z) \, dz$. This allows us to assess the extension and ionization of the disk-halo interface gas focusing only on the extraplanar gas directly below a narrow segment of the Perseus arm. These methodological differences enable a more targeted analysis of the Perseus arm's southern vertical density structure and the physical properties in the gas.

\section{Target Selection}
\label{sec:sample}

We determined the optimal region for conducting our survey, using CO \citep{Dame1987,Reid2016} and \HI\ \citep{Reid2016,Du2016} emission observations of the Perseus arm, alongside surveys of Galactic disk-halo interface \HI\ clouds \citep[e.g.,][]{Wakker2001}. We require both distances to {\it and}\ kinematics of the Perseus arm as a function of Galactic longitude to build a sample of targets where Perseus absorption can be robustly constrained. Caution must be exercised in associating ISM gas with individual spiral arms solely based on kinematics since small perturbations from Galactic rotation lead to incorrect association with distance in the outer Galaxy \citep{Burton1971,Peek2022}. Such an incorrect association of more distant gas with Perseus could lead to an incorrect estimation of its vertical extension. However, parallax-based distances to the high mass star forming regions \citep{Reid2014} and OB associations \citep{Xu2018} confirm a continuous spiral structure at the distance of the Perseus arm within $95\degree \leq l \leq 140\degree$. The detailed spiral structure, e.g., the heliocentric distance, arm width, and pitch angle, in this longitude range has been well-studied using both gas kinematics and parallax-based distances \citep{Reid2009,Reid2014,Xu2018,Xu2023,Reid2019,Laurin2024}.  Atomic and molecular line observations in the Galactic plane further show that the Perseus gas velocity range within $95\degree \leq l \leq 140\degree$ is well separated from the foreground and background gas. We, therefore, undertake our experiment at the Perseus arm segment within $95\degree \leq l \leq 140\degree$.

At these longitudes, in the Local Standard of Rest (LSR) frame, the velocity (\vlsr) of the Perseus arm material is between $-70$ to $-25$ \kms, which is well-separated from the local gas by $\sim$20--40 \kms\ \citep{Dame1987, Reid2009, Reid2014, Reid2016}. In the extraplanar region, this $[-70, -25]$ \kms\ velocity range can be associated with cloud complexes---1) infalling material accreted from the CGM; 2) material in the vertical extension of the arm; 3) galactic fountain material vertically rising from or falling into the arm  \citep{Kuntz&Danly1993,Albert2004,Lehner2022,Marasco2022}. These three types of gas structures at $|\vlsr| \simeq 40$--90 \kms\ are generally classified as IVCs \citep{Kuntz&Danly1993, Richter2001,Lehner2022} which mostly reside within $|z| \leq 1.5$ kpc \citep{Wakker2001, Lehner2022}. However, the type~1 IVCs are of external origin; therefore, if present, can contaminate the Perseus absorption. Within our chosen longitude range, the northern infalling IV Arch is one such potential contaminant to the Perseus absorption \citep{Wakker2001}. Moreover, in the second Galactic quadrant, the northern disk is warped just outside the solar circle \citep[]{Kerr1957,Oort1958,Dickey1990}, which could also impact the column density measurements. Therefore, we restrict our target search area only to the southern Galactic hemisphere. Figure~17 in \citet{Wakker2001} indicates that the PP Arch could contaminate our column density measurements at $b < -46\degree$; thus we further limit the target sample to $-46\degree < b < 0\degree$.

Since accurate column density measurements require resolving the Perseus absorption, we only consider targets observed with HST/STIS E140M, E140H, HST/COS G130M, and G160M at spectral resolutions ranging from 2.7 \kms\ (STIS E140H) to  $\sim$20 \kms\ (COS) with signal-to-noise S/N$\,\ga$ 4 per pixel. These UV spectrographs cover many species, but in this paper, we only focus on \SII, \SiIV, and \CIV\ absorption lines. As mentioned in Section \ref{sec:intro}, \CIV\ and \SiIV\ are the strongest tracers of the transition temperature gas within the STIS/COS wavelength coverage, while \SII\ probes the warm neutral and ionized gas. The two weaker transitions of the \SII\ triplet, and the \CIV\ and \SiIV\ doublets, are free of contamination from other ISM lines. Therefore, their comparison yields reliable column density estimates. We retrieved the archival spectra of the required quality for AGNs and OB-type stars within the selected longitude and latitude ranges.\footnote{Though 4 stars in our sample have the FUSE Galactic \OVI\ data, none of the AGN sightlines do. We, therefore, do not include the \OVI\ data in our analysis.} We use only OB-type stars since their stellar winds typically produce well-defined P-Cygni profiles in the high ions, ensuring that the continuum placement near the absorption lines (e.g., \CIV\ and \SiIV) can be reliably modeled \citep[e.g.,][]{Bowen2008,Lehner2011}. Their high rotation velocities also lead to broader stellar lines compared to the ISM ones, minimizing the chance of stellar photospheric features contaminating the ISM lines.

Although AGNs have cosmological distances readily placing them behind the Perseus arm, stellar distances must exceed the estimated distance of the Perseus arm in their directions. We verify this using the accurate stellar distances obtained from Gaia parallaxes \citep{Gaia}. With the release of Gaia EDR3, the reliability of parallax-based distances has improved. For stars within a few kpc of the sun, the typical parallax errors are $\sigma_{\varpi}/\varpi < 25\%$, ensuring reliable distance measurements. Using their log periodic spiral equation and the associated Perseus arm parameters from Table~2 in \citet{Reid2014}, we derive its longitude-dependent mid-arm distances and identify 6 OB-type stars, all $\geq 1\sigma$ behind the Perseus arm.

\begin{figure}[h]
   \fig{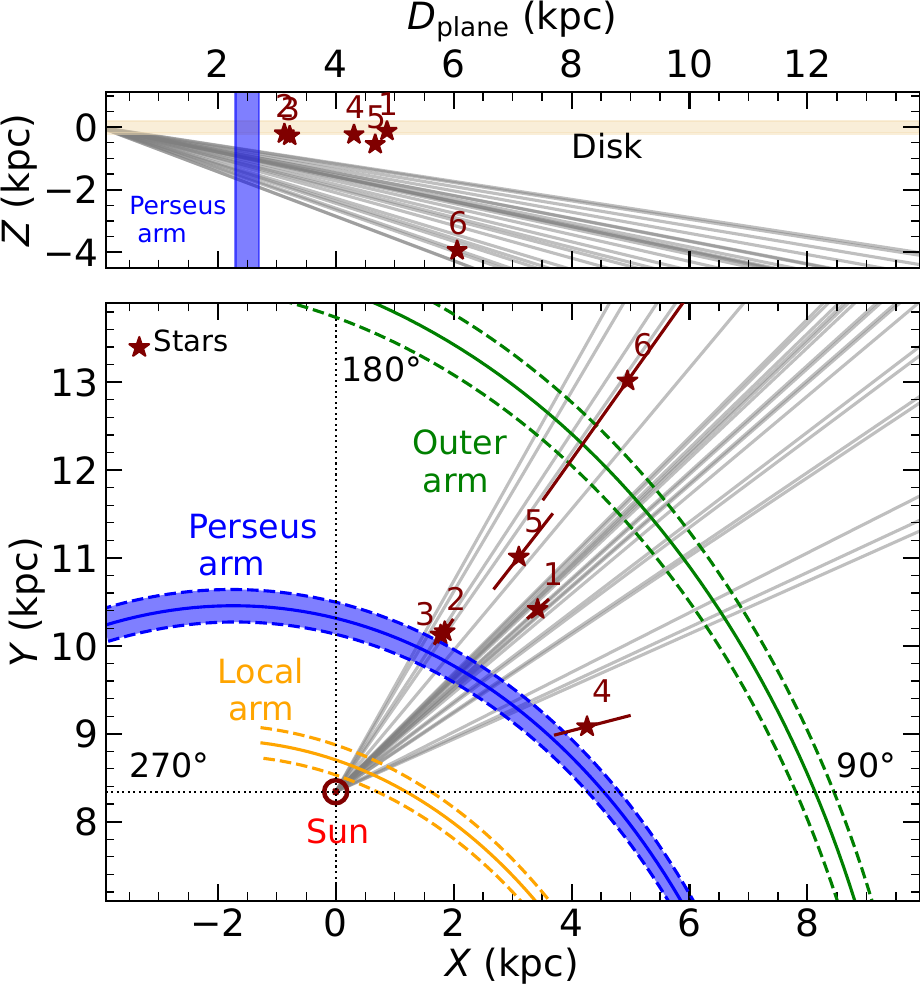}{0.95\columnwidth}{}
  \caption{The projected locations of our sightlines through the outer Galaxy. The top panel displays the $z$-projection of the sightlines as a function of radial distance on the Galactic plane, $D_{\rm plane}$, from the Sun. The thin disk, indicated by the yellow region, has a 200--300 pc scale height. Each sightline passes below the $\sim$400 pc wide Perseus arm \citep[vertical blue region,][]{Reid2014} at mid-arm height \z\ (see Table \ref{tab:sightline} Column 8). The bottom panel shows the projected in-plane positions of the Local, Perseus, and Outer spiral arms (\citealt{Reid2014}), along with our stellar and AGN sightlines that are distributed evenly over $95\degree < l < 145\degree$ and projected behind the Perseus arm.
  \label{fig:projectionplot}}
\end{figure}

Figure~\ref{fig:projectionplot} shows our sample sightline locations with respect to the Sun. The upper panel shows the projected out-of-plane positions, and the lower panel displays the projected in-plane positions of the sightlines and the spiral arms. Our sightlines are evenly distributed over a small segment of the Perseus arm within $95\degree < l < 140\degree$ and passing underneath at $0.07<|z_{\rm{P}}|<3.3$ kpc. In order of increasing $|z_{\rm{P}}|$, Table~\ref{tab:sightline} presents the target name (Column 1), target ID (Column 2), Galactic coordinates (Columns 3,4), spectral type (Column 5), parallax-based distances to the targets (Column 6), distances to the Perseus arm from \citet{Reid2014} (Column 7), sightlines' mid-arm heights, \z\ (Column 8), instruments used for target acquisition (Column 9), \SiIV\ and \CIV\ $S/N$ (Column 10).

\section{Data Origins and Analysis}
\label{sec:instrument}
In this work, we analyze the ISM $\lambda\lambda$1250, 1253 lines of the \SII\ triplet ($\lambda 1259$ line is heavily saturated and blended with the \SiII\ $\lambda 1260$ line), \SiIV\ doublet lines ($\lambda\lambda$1393, 1402), and \CIV\ doublet lines ($\lambda\lambda$1548, 1550), toward the sample stars and AGNs observed with STIS and COS spectrographs (see below). The HST data presented in this article were obtained from the Mikulski Archive for Space Telescopes (MAST) at the Space Telescope Science Institute \footnote{ The specific observations analyzed can be accessed via \dataset[doi:10.17909]{http://dx.doi.org/10.17909/35z2-az32}.}. We also use the Galactic \HI\ 21-cm emission data in the directions of our targets from the Effelsberg-Bonn \HI\ Survey \citep[EBHIS,][]{Winkel2010}. The \HI\ and \SII\ data help, in particular, constrain the Perseus velocity range (see Section \ref{subsec:velocity}). We summarize the data acquisition and reduction below.

\subsection{STIS Echelle Data}
\label{subsec:STIS}
The STIS FUV MAMA detectors took the stellar spectra in photon accumulation mode using E140H and E140M echelle gratings. Four stars (PG0122+214, HD~14434, HD~13745, BD+60~73) were observed through the $0\farcs2 \times 0\farcs2$ aperture with E140M grating, covering between 1144--1729 \AA, observing all three ions. HD~232522 was observed through the $0\farcs2 \times 0\farcs2$ aperture with E140H grating between 1140--1517 \AA, thus lacking \CIV\ coverage. For HD~210809, the $0\farcs2 \times 0\farcs09$ aperture with E140H grating covered \SII\ and the $0\farcs2 \times 0\farcs2$ aperture with E140M grating covered the high ions.

The STIS spectral resolution ranges from $R \equiv \lambda/\Delta \lambda = 114,000$ ($\Delta v = 2.7$ \kms) with E140H to $45,800$ ($\Delta v = 6.5$ \kms) with E140M. Details on STIS instrumentation can be found in \citet{woodgate1998} and \citet{Branton2021}. \citet{Kimble1998} summarizes its on-orbit performance. We retrieved the reduced 1D spectra (Pipeline \textsc{calstis} 3.4.2) from MAST. Since the \textsc{calstis} pipeline does not coadd exposures, we produced the final spectrum for each target by averaging flux from each exposure and overlapping spectral orders, weighted by the flux error. We shift the final coadded spectra from the heliocentric wavelength scale to the LSR frame.

\subsection{COS Data}
\label{subsec:COS}
All the AGNs in our sample are from Project AMIGA \citep{Lehner2020}, and we adopted their reduced coadded COS G130M and G160M spectra, providing a continuous wavelength coverage between 1140--1770 \AA. We refer the reader to Section~2 in \citet{Lehner2020} for more details on the alignment and coaddition procedures of these data. The important outcome from their careful alignments is that the absolute COS velocity calibration is accurate to better than a few \kms.

\subsection{\HI\ 21-cm Data}
\label{subsec:HI}

We use \HI\ 21-cm emission data of the sky toward each target from the EBHIS survey \citep{Winkel2010}. The \HI\ 21-cm emission in these directions has been mapped with the Effelsberg 100-meter Radio Telescope at angular resolution $ \sim 10\farcm8$ and velocity resolution $\Delta v \sim 1.44$ \kms. The high angular resolution of the EBHIS data is advantageous for comparing the \HI\ 21-cm profiles against the STIS \SII\ spectra. However, some of the kinematic differences between the UV absorption and \HI\ emission data occur owing to comparing essentially a pencil-beam sightline to a $ \sim 10.8\arcmin$ pointing and the \HI\ emission arising both in the foreground and background of the stars. The EBHIS data reduction pipeline applies stray radiation correction, RFI subtraction, baseline correction, flux, and wavelength calibration. Details of the reduction process and data products can be found in \citet{Winkel2010}.

\begin{figure*}
\fig{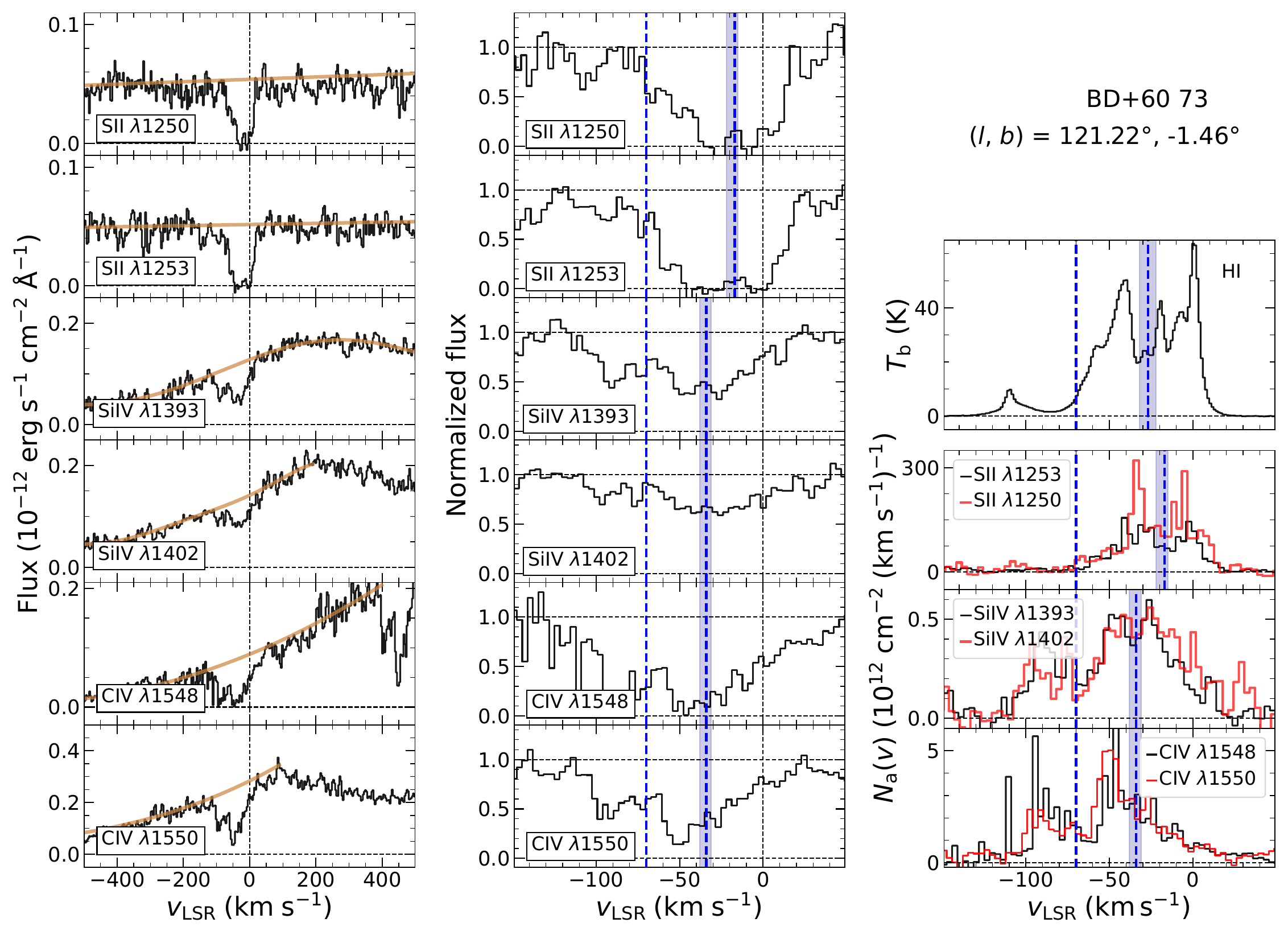}{0.95\textwidth}{}
\caption{{\it Left}: the observed flux and continuum models (solid golden lines) and {\it middle}: normalized flux of \SII\ ($\lambda\lambda$1250, 1253), \SiIV\ ($\lambda\lambda$1393, 1402), and \CIV\  ($\lambda\lambda$1548, 1550) against the LSR velocity toward BD+60~73.  {\it Right}: the \HI\ 21-cm brightness temperature profile, $T_b(v)$, and apparent column density, \Nav, profiles derived from the normalized flux of \SII, \SiIV, and \CIV\ transitions. In the left and middle panels, we mark the absorption from the local gas centered at $v=0$ \kms\ by the black-dash line. In the middle and right panels, the blue-dash lines indicate the velocity range [$v_1$, $v_2$] of the absorption from the Perseus arm material, and the blue shaded region around $v_2$ indicates the corresponding systematic uncertainty (see Section \ref{subsec:velocity} for more details). The other sightlines are presented in the Appendix.}
\label{fig:BD+6073}
\end{figure*}

\subsection{Continuum Normalization and Column Density Profiles}
\label{subsec:continuum}

To analyze the STIS and COS spectra, we model the continuum flux near the ISM absorption lines, which involves two main steps. First, we select regions on either side of the ISM absorption for continuum fitting. These regions span roughly 200--1000 \kms\ in the AGN spectra, and 50--400 \kms\ in the stellar spectra. The exact velocity range depends on the continuum properties near each ISM transition. Second, we fit the continuum flux in the selected regions near the ISM absorption with Legendre polynomials of varying orders \citep{Sembach1992}. AGN continua are typically straightforward and fitted with a 1st or 2nd-order polynomial. Stellar continua are intrinsically more complex due to complicated wind profiles and photospheric lines, requiring a 1st to 5th-order fit. The continuum normalized flux is defined as $F_n(v) = F_o(v)/F_c(v)$, where $F_o(v)$ is the observed flux and $F_c(v)$ is the continuum model. We then convert $F_n(v)$ to apparent optical depth $\tau_a(v) = -\ln[F_n(v)]$ and derive the apparent column density, $N_a(v)$, using $N_a(v) = 3.768 \times 10^{14} \, \tau_a(v)/(f\lambda)$ ${\rm cm}^{-2}\,(\kms)^{-1}$ \citep{Savage1991}, where $f$ is the oscillator strength and $\lambda$ is the wavelength in \AA. The \Nav\ profile of each transition is used for the rest of the analysis.

Figure~\ref{fig:BD+6073} shows an example---the raw flux with the continuum fits (left panels) and normalized flux (middle panels) of the \SII\ $\lambda\lambda$1250, 1253 transitions, \SiIV, and \CIV\ doublets toward BD+60~73. The other sightlines are presented in the Appendix \ref{sec:app-A}. As we see, the continua around all the ISM lines toward BD+60~73 are straightforward. However, for two stars, owing to the unreliable continuum model, we could use only one of the \SiIV\ transitions---$\lambda$1402 line toward HD~210809 and \SiIV\ $\lambda$1393 toward HD~13745.

In the right panel of Figure~\ref{fig:BD+6073}, we show the \HI\ 21-cm brightness temperature profile, $T_b (v)$ (K), in the direction of BD+60~73 at the top, followed by the $N_a(v)$ profiles of \SII, \SiIV, \CIV\ lines. This aids in understanding the evolution of the emission/absorption line profiles with changing ionization energy, thus confidently identifying the Perseus velocity range in each ion and \HI\ (see Section \ref{subsec:velocity}).

In each bottom right panel, overplotting an ion's $N_a(v)$ profiles of two transitions with different strengths ($f \lambda$) also helps diagnose any potential contamination or saturation issues. The former may affect any part of a line, while saturation predominantly affects the line core. Section~\ref{subsection:column} describes our saturation treatment. Toward our stars, the ISM lines are free of contamination from circumstellar gas, wind, or photospheric absorption. In the case of the AGNs, higher redshift absorbers can contaminate the ISM absorption, as evident in several cases. We removed the following contaminated transitions from our sample: \SiIV\ $\lambda$1402 toward PG0003+158, \SII\ $\lambda$1250 and \CIV\ $\lambda$1548 toward 4C25.01, \CIV\ $\lambda$1550 toward MRK335 and 3C66A, and \SII\ $\lambda$1250 toward Zw535.012.

\subsection{The Perseus Velocity Range}
\label{subsec:velocity}

To delineate the velocity range associated with Perseus absorption/emission, we first inspect the \HI\ and \SII\ profiles, which have narrower and better-defined components than the high ions. However, the larger beam size and broader intrinsic line width of \HI, as well as \SII\ arising in the ionized (rather than neutral) regions of the gas, may lead to slight differences between \HI\ and \SII\ velocity ranges. Additionally, \HI\ may include emission from the material behind the stars and Perseus arm. We, therefore, first define the Perseus velocity range using \SII\ and cross-check it against \HI\ before assessing \CIV\ and \SiIV.

Visually, two major components are identifiable in \SII\ and \HI: one component centered at $v \sim 0$ \kms\ associated with the local gas, and the other component centered at $-45\pm 5$ \kms\ associated with the Perseus arm. The high ion Perseus absorption may be at more negative velocities, e.g., shifted by $-15$ to $-20$ \kms\ relative to \SII\ and \HI\ toward HD~14434, HD~13745 and HD~232522. These stars are situated in regions near intense stellar activity in the Perseus arm \citep{Knauth2003, Cappa2000}. Consequently, the highly blue-shifted high ion Perseus absorption compared to \SII\ in these sightlines is possibly associated with expanding stellar windblown bubbles in the Perseus arm \citep{Knauth2003}.

When Perseus and local gas are resolved in \SII, we define $v_2$, Perseus velocity upper bound, at the shoulder between components where the \Nav\ profile intensity drops significantly, i.e., the local minima of the \Nav\ profiles between Perseus and local absorption (at $\sim 17$ \kms\ for the example in Figure \ref{fig:BD+6073}). The adopted \SII\ $v_2$ generally agrees with \HI; where discrepant, we define \HI\ $v_2$ separately using the same approach. For the blended components, clues from both \SII\ and \HI\ set $v_2\approx -25\pm10$ \kms. When local and Perseus absorption components are resolved in the high ion profiles, we use the same technique to define $v_2$ in \SiIV\ and \CIV; otherwise, we adopt $v_2$ from \SII. We choose a nominal spread of 2--5 \kms\ about $v_2$ to account for the systematic uncertainties associated with separating Perseus and local gas in each ion.

Careful attention sets $v_1$, the Perseus velocity lower bound, since \HI\ and AGN sightlines may show absorption from the material behind the Perseus arm at $\vlsr<-90$ \kms. However, no sightline shows \SII\ absorption at $\vlsr<-90$ \kms. \SII\ $v_1$ is set where the \Nav\ profile first drops to zero (usually $-90<v_1<-50$ \kms) and is adopted for the other ions and \HI. Exceptions are the HD~14434, HD~13745, and HD~232522 sightlines, where high and low ions' velocity ranges are very different, and toward HD~210809, where the high ion Perseus absorption is broader and extends to more negative velocity. Thus, in these sightlines, high ion $v_1$ is more negative than that of \SII, but in all cases $v_1 \geq -100$ \kms.

Figure \ref{fig:BD+6073} shows the adopted Perseus velocity ranges $[v_1,v_2]$ and the systematic uncertainties on $v_2$ (the blue-dash lines and the blue shaded region in the middle and right panels) toward BD+60~73. Here, the same $v_1$ is adopted for all the ions and \HI. The Perseus and local absorption components are resolved in \HI, \SII, and \SiIV, and thus, their $v_2$ are set at the trough between the Perseus and local absorption. For \CIV, we adopt \SiIV\ $v_2$. In this direction, the high ion $v_2$ is 12 \kms\ lower than the \SII\ $v_{2}$. Table \ref{tab:vel-range} compiles the adopted $[v_1,v_2]$ and systematic uncertainties on $v_2$ in each ion along each direction.

\begin{deluxetable*}{lCCCC}
\tablenum{2} \tabletypesize{\footnotesize}
\tablecaption{Perseus Velocity Integration Ranges, [$v_1, v_2$]. 
\label{tab:vel-range}}
\tablehead{\colhead{Targets} & \colhead{H I} & \colhead{S II } & \colhead{C IV } & \colhead{Si IV}\\ \colhead{ } & \colhead{(km s$^{-1}$)} & \colhead{(km s$^{-1})$} & \colhead{(km s$^{-1}$) } & \colhead{(km s$^{-1}$)}}
\startdata
BD+60~73 & $[-70 ,\; -27\pm5]$ & $[-70 ,\; -17 \; (^{+2}_{-5})]$ & $[-70,\; -34 \; (^{+3}_{-4})]$ & $[-70,\; -34 \; (^{+3}_{-4})]$ \\
HD~14434 & $[-73 ,\; -22\pm2]$ & $[-73 ,\; -32\pm2]$ & $[-100,\; -30\pm5]$ & $[-100,\; -30\pm5]$ \\
HD~13745 & $[-85 ,\; -25 \; (^{+2}_{-4})]$ & $[-85 ,\; -25 \; (^{+2}_{-4})]$ & $[-100,\; -25 \; (^{+3}_{-2})]$ & $[-100,\; -25 \; (^{+3}_{-2})]$ \\
HD~210809 & $[-50 ,\; -20\pm5]$ & $[-50 ,\; -20\pm5]$ & $[-70,\; -18\pm3]$ & $[-70,\; -18\pm3]$ \\
HD~232522 & $[-70 ,\; -26\pm2]$ & $[-70 ,\; -26\pm2]$ & $\nodata ^a$ & $[-90,\; -35\pm3]$ \\
3C66A & $[-80 ,\; -20\pm5]$ & $[ -80 ,\;  -20\pm5]$ & $[-80 ,\; -20\pm5]$ & $[ -80 ,\;  -20\pm5]$ \\
Zw535.012 & $[-80 ,\; -42 \; (^{+4}_{-3})]$ & $[ -80 ,\;  -35\pm5]$ & $[-80 ,\; -35\pm5]$ & $[ -80 ,\;  -35\pm5]$ \\
FBS0150+396 & $[-70 ,\; -18 \; (^{+3}_{-4})]$ & $[-70 ,\; -18 \; (^{+3}_{-4})]$ & $[-70 ,\; -18 \; (^{+3}_{-4})]$ & $[ -70 ,\;  -18\pm3]$ 
\\
IRASF00040+4325 & $[-80 ,\; -20\pm1]$ & $[ -80 ,\;  -20\pm1]$ & $[-80 ,\; -20\pm1]$ & $[ -80 ,\;  -20\pm1]$ \\
UVQSJ001903.85+423809.0 & $[-80 ,\; -24\pm2]$ & $[ -75 ,\;  -24\pm2]$ & $[-75 ,\; -24\pm2]$ & $[ -75 ,\;  -24\pm2]$ \\
PGC2304 & $[-75 ,\; -25\pm5]$ & $[ -75 ,\;  -25\pm2]$ & $[-75 ,\; -25\pm2]$ & $[ -75 ,\;  -25\pm2]$ \\
RXSJ0118.8+3836 & $[-80 ,\; -15\pm1]$ & $[ -80 ,\;  -15\pm1]$ & $[-80 ,\; -15\pm1]$ & $[ -80 ,\;  -15\pm1]$ \\
IVZw29 & $[-80 ,\; -20\pm2]$ & $[ -80 ,\;  -20\pm2]$ & $[-80 ,\; -20\pm2]$ & $[ -80 ,\;  -20\pm2]$ \\
RXJ0049.8+3931 & $[-75 ,\; -20 \; (^{+5}_{-3})]$ & $[-75 ,\; -20 \; (^{+5}_{-3})]$ & $[-75 ,\; -20 \; (^{+5}_{-3})]$ & $[ -75 ,\;  -20\pm5]$ \\
RXJ0048.3+3941 & $[-75 ,\; -20\pm1]$ & $[ -75 ,\;  -20\pm1]$ & $[-75 ,\; -20\pm1]$ & $[ -75 ,\;  -20\pm1]$ \\
LAMOSTJ003432.52+391836.1 & $[-75 ,\; -25\pm2]$ & $[ -75 ,\;  -25\pm2]$ & $[-75 ,\; -25\pm2]$ & $[ -75 ,\;  -25\pm2]$ \\
RXSJ0155.6+3115 & $[-70 ,\; -25 \; (^{+3}_{-5})]$ & $[ -70 ,\;  -20\pm1]$ & $[-70 ,\; -20\pm1]$ & $[ -70 ,\;  -20\pm1]$ \\
RXJ0043.6+3725 & $[-70 ,\; -20 \; (^{+4}_{-5})]$ & $[ -70 ,\;  -20\pm1]$ & $[-70 ,\; -20\pm1]$ & $[ -70 ,\;  -20\pm1]$ \\
3C48.0 & $[-75 ,\; -27 \; (^{+2}_{-3})]$ & $[-75 ,\; -27 \; (^{+2}_{-3})]$ & $[-75 ,\; -27 \; (^{+2}_{-3})]$ & $[ -75 ,\;  -27\pm2]$ \\
RXJ0050.8+3536 & $[-75 ,\; -20\pm5]$ & $[ -75 ,\;  -20\pm1]$ & $[-75 ,\; -20\pm1]$ & $[ -75 ,\;  -20\pm1]$ \\
RXJ0028.1+3103 & $[-75 ,\; -32 \; (^{+2}_{-3})]$ & $[ -75 ,\;  -25\pm5]$ & $[-75 ,\; -25\pm5]$ & $[ -75 ,\;  -25\pm5]$ \\
PG0052+251 & $[-66 ,\; -20 \; (^{+1}_{-2})]$ & $[-66 ,\; -20 \; (^{+1}_{-2})]$ & $[-66 ,\; -20 \; (^{+1}_{-2})]$ & $[ -66 ,\;  -20\pm1]$ \\
PG0122+214 & $[-70 ,\; -20 \; (^{+1}_{-3})]$ & $[-70 ,\; -20 \; (^{+1}_{-3})]$ & $[-70,\; -20 \; (^{+5}_{-5})]$ & $[-80,\; -20\pm5]$ \\
4C25.01 & $[-70 ,\; -20 \; (^{+1}_{-4})]$ & $[-70 ,\; -20 \; (^{+1}_{-4})]$ & $[-75 ,\; -20 \; (^{+1}_{-4})]$ & $[ -70 ,\;  -20\pm1]$ \\
RXJ0053.7+2232 & $[-75 ,\; -20\pm1]$ & $[ -75 ,\;  -20\pm1]$ & $[-75 ,\; -20\pm1]$ & $[ -75 ,\;  -20\pm1]$ \\
RBS2055 & $[-70 ,\; -22\pm2]$ & $[ -70 ,\;  -22\pm2]$ & $[-70 ,\; -22\pm2]$ & $[ -70 ,\;  -22\pm2]$ \\
MRK1148 & $[-65 ,\; -21 \; (^{+1}_{-4})]$ & $[ -65 ,\;  -25\pm5]$ & $[-75 ,\; -25\pm5]$ & $[ -65 ,\;  -25\pm5]$ \\
MRK335 & $[-90 ,\; -21 \; (^{+2}_{-3})]$ & $[ -90 ,\;  -30\pm2]$ & $[-90 ,\; -30\pm2]$ & $[ -90 ,\;  -30\pm2]$ \\
PG0003+158 & $[-70 ,\; -22 \; (^{+2}_{-3})]$ & $[ -70 ,\;  -25\pm5]$ & $[-70 ,\; -25\pm5]$ & $[ -70 ,\;  -25\pm5]$\\
\enddata
\tablecomments{ The errors represent the systematic uncertainties on $v_2$. $^a$ no \CIV\ coverage toward HD~232522.}
\end{deluxetable*}

\subsection{The Total Apparent Column Density of Perseus Gas}
\label{subsection:column}

For each ion, we derive the total apparent Perseus column density, $N_a$, of each transition by integrating its $N_a (v)$ profile over the Perseus velocities $[v_1,v_2]$, e.g., $N_a = \int_{v_1}^{v_2} N_a (v) dv$ \citep{Savage1991}. We report the weighted average as the adopted Perseus column density, $N(z_{\rm P})$, of an ion when its two transitions are unsaturated such that the apparent column densities, $N_a$, estimated for the two transitions of the ion agree within $1\sigma$. All \SiIV\ and \CIV\ doublets in our sample are unsaturated. The \SiIV\ and \CIV\ in two sightlines are not detected at $3 \sigma$ significance, we quote $3 \sigma$ upper limit on the column densities defined as $3\times$ the $1 \sigma$ error derived for the column density (integrated over the velocity reported in Table \ref{tab:vel-range}) assuming that the absorption line lies on the linear part of the curve of growth. For \HI, we report the adopted Perseus column density, $N(z_{\rm P}) = N_a = 1.8\times10^{18} \;{\rm cm^{-2}}\int_{v_1}^{v_2}T_b(v) dv$, where $T_b(v)$ (in K) is the \HI\ 21 cm brightness temperature profile.

In \SII, when there is evidence of saturation and $\Delta\log{N_a}$ (the difference between the weak and strong line column densities) of the two \SII\ lines is $\leq0.12$ dex, we apply a saturation correction to the column density following \citet{Savage1991} and \citet{Wotta2016}. Larger differences indicate saturation is too significant to correct. In that case, we report the \SII\ $\lambda1253$ line's $N_a$ value as a lower limit on $N(z_{\rm P})$.

We also estimate the systematic uncertainty in $N_a$ caused by the possible variations in the velocity range used for integrating the \Nav\ profiles---related to the systematic uncertainty on $v_2$ (see Section \ref{subsec:velocity}). Table~\ref{tab:column} presents the adopted column densities, $N(z_{\rm P})$, with statistical and systematic uncertainties or lower/upper limits in cases of saturation/non-detection.

\section{Results}
\label{sec:Results}

\begin{figure*}
   \fig{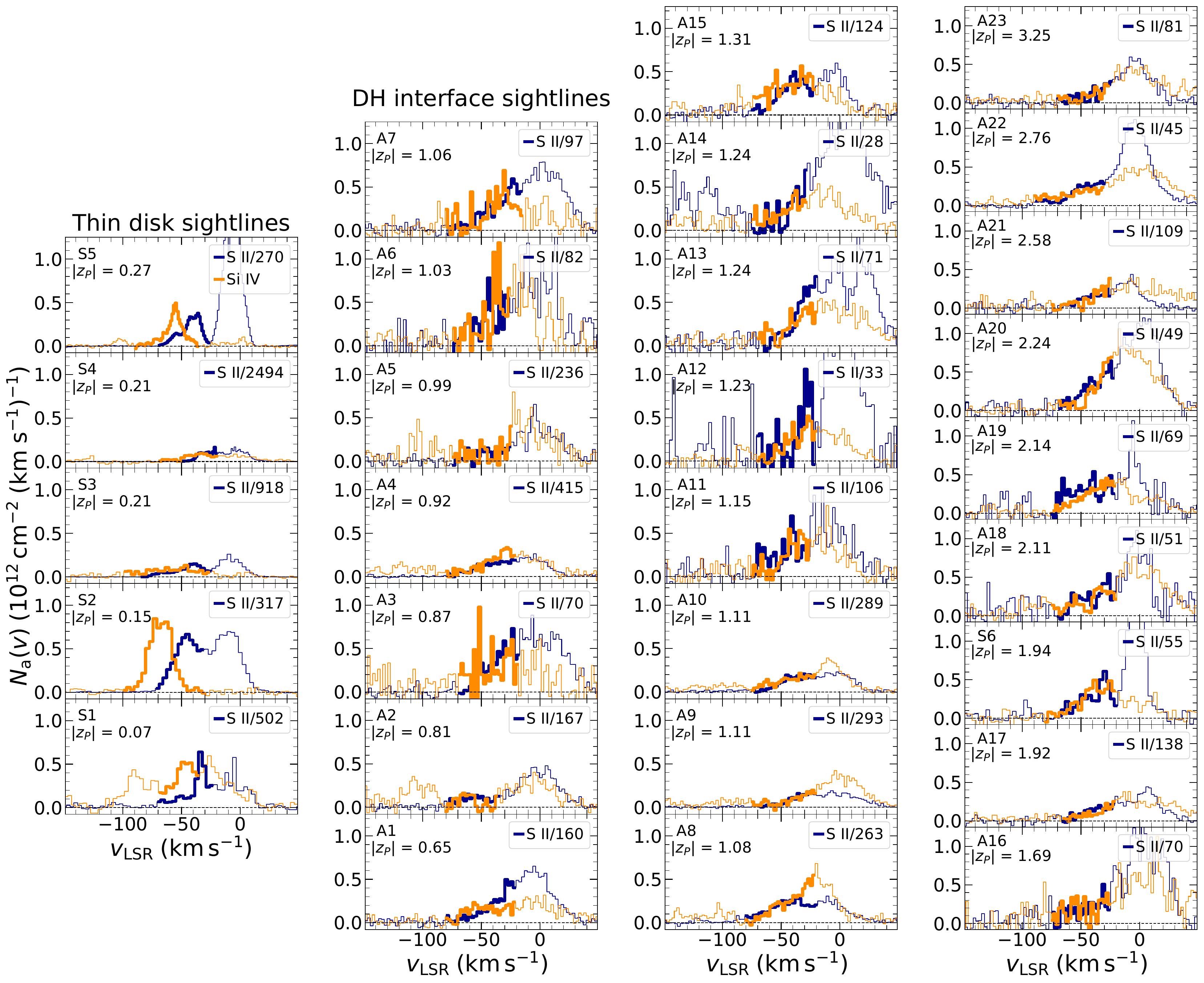}{0.9\textwidth}{}
  \caption{Toward each sightline, we compare the \SiIV\ \Nav\ (orange) profile to the scaled \SII\ profile (blue). The \SII\ scaling factor in each sightline is the average \siIVsII\ value estimated within the Perseus velocity range of that sightline. The sightline ID, mid-arm height \z\ (kpc) below the Perseus arm are given on the left, and the \SII\ scaling factor is given on the right of each panel. The Perseus absorption toward each sightline in each ion is highlighted with a thicker line. The leftmost panels are thin-disk sightlines at $|z_{\rm{P}}|<0.5$ kpc below Perseus arm, and the rest are the disk-halo interface sightlines at $|z_{\rm{P}}|>0.5$ kpc. The sightlines are stacked from the bottom to the top in ascending mid-arm heights, \z.
  \label{fig:scale-SII-SiIV}}
\end{figure*}

\subsection{Comparison of the \Nav\ Profiles}
\label{subsec:Nav-comparison}

In Figure \ref{fig:scale-SII-SiIV}, we compare the \SiIV\ \Nav\ profiles to the \SII\ profiles. For visual comparison, toward each sightline, we scale the \SII\ profile down by the \siIVsIIav\ value estimated over the Perseus velocity range along the sightline. We only focus on the Perseus absorption in each profile and find: 1) large scaling factors are needed for \SII\ profiles at all \z, indicating that \SII\ profiles are significantly stronger than \SiIV; 2) in the thin disk, the scaled \SII\ \Nav\ profiles are usually kinematically very different from the \SiIV\ profiles over the Perseus velocity range; 3) in the disk-halo interface, \SiIV\ and scaled \SII\ \Nav\ profiles kinematically follow each other very well. Additionally, in the thin disk, the \SiIV\ to \SII\ column density ratios vary substantially at different velocities. In the disk-halo interface, column density ratios tend to be very similar at all Perseus velocities toward a given sight line---consequence of good kinematic agreement between the \SII\ and \SiIV\ profiles in the disk-halo interface. The well-matched scaled \SII\ and \SiIV\ \Nav\ profiles in the halo star PG0122+214's medium-resolution STIS spectra indicates that the agreement in \SII\ and \SiIV\ COS AGN spectra is not an artifact from the coarser COS resolution. The disk-halo interface sightlines, therefore, suggest a physical relationship between \SII\ and \SiIV\ in the extraplanar Perseus gas, which is not the case for the thin disk sightlines.

\begin{figure*}
   \fig{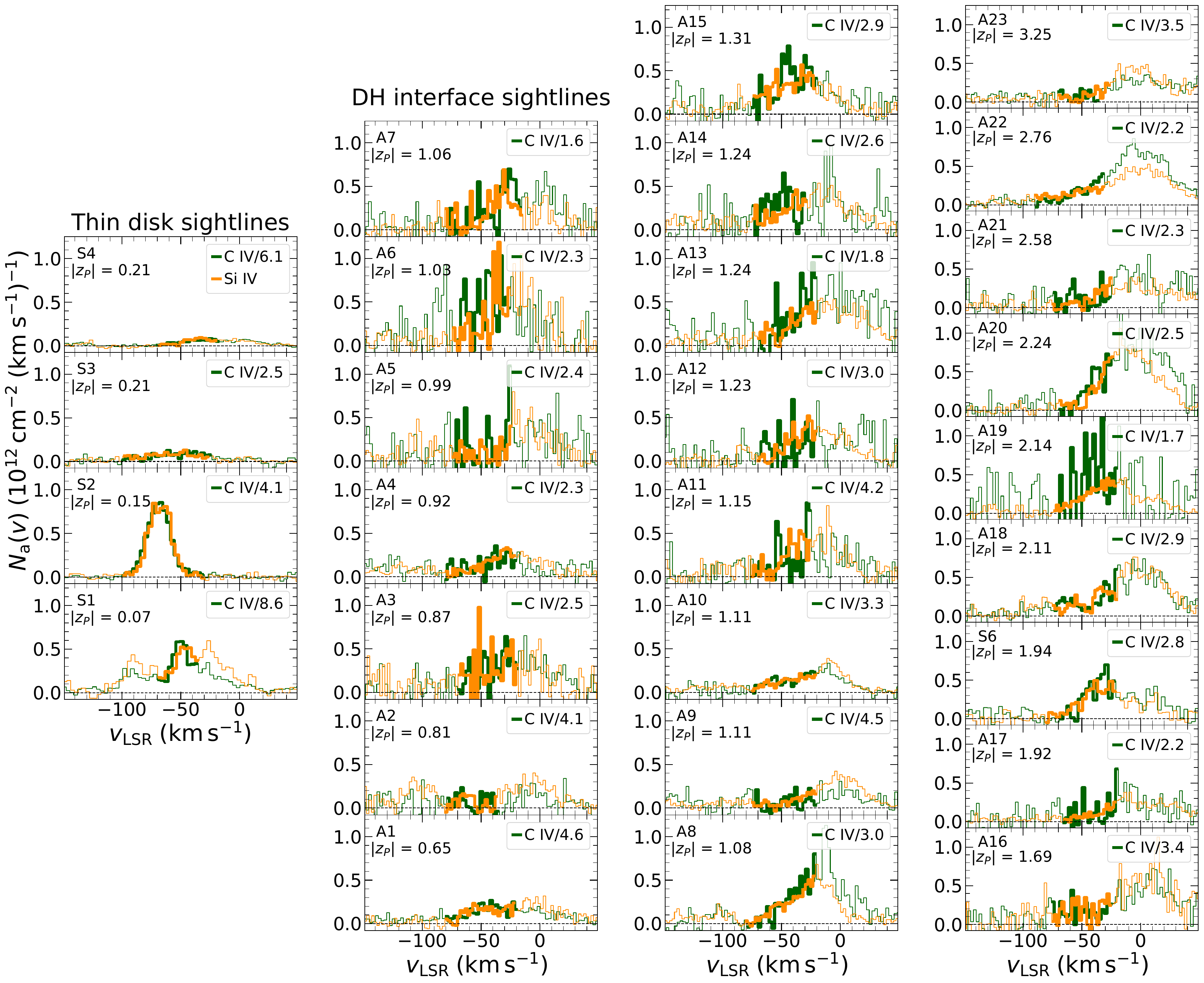}{0.9\textwidth}{}
  \caption{Similar to \ref{fig:scale-SII-SiIV}, here we compare the \SiIV\ (orange) \Nav\ profile to the \CIV\ (green) profile, which is scaled down by the average \cIVsiIV\ value estimated over the Perseus velocity range of each sightline.
  \label{fig:scale-CIV-SiIV}}
\end{figure*}

In Figure \ref{fig:scale-CIV-SiIV}, we compare \SiIV\ and \CIV\ \Nav\ profiles. In this case, toward each sightline, the \CIV\ profile is scaled by the \cIVsiIVav\ value estimated over Perseus velocities along the sightline. Based on this comparison, we find: 1) \CIV\ is usually 2--4 times stronger than \SiIV, except for two thin disk stars, HD~210809 and BD+60~73, where \CIV\ profiles are $>5 \times$ stronger than \SiIV; 2) either ion's \Nav\ profiles are equally strong in the thin disk and the disk-halo interface; and 3) \CIV\ and \SiIV\ \Nav\ profiles closely mirror each other in all the sightlines, indicating these two ions' kinematic similarities in the thin disk and disk-halo interface. Additionally, the ratio of \CIV\ to \SiIV\ column densities tends to be similar at all Perseus velocities in the thin disk and the disk-halo interface---consequence of the good kinematic agreement between the \Nav\ profiles of these two ions. This suggests that \CIV\ and \SiIV\ are physically correlated both in the thin disk and the disk-halo interface gas sampled by our sightlines. This also suggests co-spatiality between \SII\ and \CIV\ bearing gas only in the disk-halo interface below the Perseus arm (see Appendix \ref{sec:app-B}).

\begin{deluxetable*}{lCCCC}
\tablenum{3} 
\tabletypesize{\footnotesize}
\tablecaption{Adopted Perseus Column Densities.}
\tablehead{ \colhead{Target} & \colhead{$\log N_{\rm {H\,I}}\,(z_{\rm P})$} & \colhead{$\log N_{\rm {S\,II}}\, (z_{\rm P})$} & \colhead{ $\log N_{\rm {Si\,IV}}\, (z_{\rm P})$} & \colhead{$\log N_{\rm {C\,IV}}\, (z_{\rm P})$}\\ 
\colhead{ } & \colhead{$[\rm {cm}^{-2}]$} &  \colhead{$[\rm{cm}^{-2}]$} & \colhead{$[\rm{cm}^{-2}]$} & \colhead{$[\rm{cm}^{-2}]$} }
\startdata
BD+60~73 & $21.321 \pm 0.001 \; \
                (^{+0.042}_{-0.034})$ & $>15.797 $& $13.096 \pm 0.022 \; (^{+0.041}_{-0.059})$ & $14.028 \pm 0.021 \; (^{+0.024}_{-0.045})$ \\
HD~14434 & $21.225 \pm 0.001 \; \
                (^{+0.013}_{-0.007})$ & $>15.833$ & $13.331 \pm 0.010 \; \pm 0.002$ &$ 13.944 \pm 0.006 \; (^{+0.003}_{-0.005})$ \\
HD~13745 & $21.105 \pm 0.001 \; \
                (^{+0.016}_{-0.028})$ & $15.671 \pm 0.033
                $ & $12.708 \pm 0.056 \; (^{+0.016}_{-0.009})$ & $13.110 \pm 0.055 \; (^{+0.032}_{-0.038})$ \\
                HD~210809 & $20.998 \pm 0.001 \; \
                (^{+0.045}_{-0.057})$ & $>15.807$ & $12.410 \pm 0.053 \; \pm 0.032$ & $13.194 \pm 0.018 \; (^{+0.068}_{-0.041})$ \\
HD~232522 & $20.648 \pm 0.003 \; \
                (^{+0.022}_{-0.012})$ & $15.278 \pm 0.003 \; \
                (^{+0.006}_{-0.005})$ & $12.846 \pm 0.018 \; (^{+0.001}_{-0.002})$ & $\nodata$ \\
3C66A & $20.302 \pm 0.008 \; \
                (^{+0.103}_{-0.140})$ & $15.140 \pm 0.020 \; \
                (^{+0.090}_{-0.100})$ & $12.935 \pm 0.034 \; (^{+0.038}_{-0.048})$ & $13.599 \pm 0.037 \; (^{+0.037}_{-0.055})$\\
Zw535.012 & $19.718 \pm 0.023 \; \
                (^{+0.047}_{-0.044}) $& $14.676 \pm 0.034 \; \
                (^{+0.060}_{-0.063})$ & $<12.452$ & $<13.062$ \\
FBS0150+396 & $20.086 \pm 0.012 \; \
                (^{+0.048}_{-0.065})$ & $15.014 \pm 0.037 \; \
                (^{+0.055}_{-0.050})$ & $13.163 \pm 0.091 \; (^{+0.013}_{-0.025})$ & $13.560 \pm 0.040 \; \pm 0.030 $\\
IRASF00040+4325 & $20.300 \pm 0.008 \; \
                (^{+0.013}_{-0.015})$ & $>15.552$ & $12.933 \pm 0.028 \; \pm 0.013$ & $13.288 \pm 0.053 \; (^{+0.012}_{-0.010})$ \\
UVQSJ001903.85+423809.0 & $20.177 \pm 0.011 \; \
                (^{+0.023}_{-0.054})$ & $>15.080$ & $<12.706$ & $<13.090$ \\
PGC2304 & $20.113 \pm 0.012 \; \
                (^{+0.061}_{-0.071})$ & $15.090 \pm 0.050 \; \
                (^{+0.020}_{-0.030})$ & $13.175 \pm 0.148 \; (^{+0.032}_{-0.040})$ & $13.540 \pm 0.070 \; \pm 0.020$ \\
RXSJ0118.8+3836 & $20.177 \pm 0.010 \; \
                (^{+0.017}_{-0.016})$ & $15.110 \pm 0.020 \; \
                \pm 0.020$ & $13.121 \pm 0.071 \; (^{+0.003}_{-0.006})$ & $13.331 \pm 0.094 \; (^{+0.012}_{-0.010})$ \\
IVZw29 & $20.371 \pm 0.006 \; \
                (^{+0.009}_{-0.021})$ & $>15.565$ & $13.145 \pm 0.026 \; (^{+0.037}_{-0.039})$ & $13.620 \pm 0.020 \; (^{+0.030}_{-0.040})$ \\
RXJ0049.8+3931 & $20.236 \pm 0.008 \; \
                (^{+0.047}_{-0.024})$ & $15.123 \pm 0.020 \; \
                (^{+0.037}_{-0.079})$ & $12.656 \pm 0.042 \; (^{+0.100}_{-0.052})$ & $13.310 \pm 0.040 \; (^{+0.070}_{-0.030})$ \\
RXJ0048.3+3941 & $20.212 \pm 0.008 \; \
                (^{+0.011}_{-0.012})$ & $15.287 \pm 0.027 $ & $12.826 \pm 0.023 \; (^{+0.014}_{-0.013})$ & $13.350 \pm 0.030 \; \pm 0.010$ \\
LAMOSTJ003432.52+391836.1 & $19.998 \pm 0.015 \; \pm 0.021$ & $15.060 \pm 0.040 \; \
                \pm 0.030$ & $13.031 \pm 0.065 \; (^{+0.028}_{-0.032})$ & $13.650 \pm 0.090 \; (^{+0.030}_{-0.040})$ \\
RXSJ0155.6+3115 & $19.429 \pm 0.044 \; \
                (^{+0.066}_{-0.121})$ & $14.530 \pm 0.080 \; \
                (^{+0.030}_{-0.020})$ & $13.003 \pm 0.049 \; (^{+0.016}_{-0.015})$ & $13.477 \pm 0.078 \; \pm 0.010$ \\
RXJ0043.6+3725 & $20.167 \pm 0.010 \; \
                (^{+0.028}_{-0.045})$ & $14.930 \pm 0.030 \; \
                \pm 0.020$ & $13.073 \pm 0.040 \; \pm 0.015$ & $13.340 \pm 0.060 \; \pm 0.020$\\
3C48.0 & $19.422 \pm 0.064 \; \
                (^{+0.059}_{-0.057})$ & $14.436 \pm 0.080 \; \
                (^{+0.061}_{-0.110})$ & $12.981 \pm 0.046 \; (^{+0.030}_{-0.045})$ & $13.390 \pm 0.060 \; (^{+0.030}_{-0.040}) $\\
RXJ0050.8+3536 & $20.110 \pm 0.012 \; \
                (^{+0.043}_{-0.047})$ & $>15.349$ & $13.253 \pm 0.031 \; \pm 0.006$ & $13.720 \pm 0.030 \; \pm 0.010$\\
RXJ0028.1+3103 & $19.422 \pm 0.048 \; \
                (^{+0.021}_{-0.060})$ & $14.800 \pm 0.080 \; \
                (^{+0.132}_{-0.057})$ & $12.950 \pm 0.109 \; (^{+0.064}_{-0.085})$ & $13.480 \pm 0.040 \; \pm 0.050$ \\
PG0052+251 & $19.601 \pm 0.034 \; \
                (^{+0.018}_{-0.040})$ & $14.830 \pm 0.020 \; \
                (^{+0.020}_{-0.040})$ & $12.688 \pm 0.032 \; (^{+0.025}_{-0.055})$ & $13.041 \pm 0.077 \; (^{+0.025}_{-0.046})$ \\
PG0122+214 & $19.464 \pm 0.052 \; \
                (^{+0.022}_{-0.039})$ & $14.872 \pm 0.021 \; \
                (^{+0.008}_{-0.025})$ & $13.131 \pm 0.037 \; (^{+0.053}_{-0.066})$ & $13.573 \pm 0.037 \; (^{+0.047}_{-0.070})$ \\
4C25.01 & $19.396 \pm 0.054 \; \
                (^{+0.024}_{-0.065})$ & $14.750 \pm 0.071 \; \
                (^{+0.020}_{-0.047})$ & $13.035 \pm 0.049 \; (^{+0.017}_{-0.051})$ & $13.493 \pm 0.052 \; (^{+0.025}_{-0.078})$ \\
RXJ0053.7+2232 & $19.590 \pm 0.036 \; \pm 0.013$ & $14.940 \pm 0.040 \; \
                \pm 0.010$ & $13.101 \pm 0.027 \; \pm 0.013$ & $13.333 \pm 0.114 \; (^{+0.018}_{-0.019})$ \\
RBS2055 & $19.681 \pm 0.025 \; \
                (^{+0.037}_{-0.018})$ & $14.770 \pm 0.020 \; \
                \pm 0.030$ & $13.078 \pm 0.041 \; (^{+0.052}_{-0.059})$ & $13.470 \pm 0.040 \; \pm 0.050$ \\
MRK1148 & $19.370 \pm 0.054 \; \
                (^{+0.013}_{-0.051})$ & $14.740 \pm 0.030 \; \
                (^{+0.060}_{-0.090})$ & $12.702 \pm 0.073 \; (^{+0.093}_{-0.136})$ & $13.069 \pm 0.090 \; (^{+0.123}_{-0.113})$ \\
MRK335 & $19.551 \pm 0.046 \; \
                (^{+0.028}_{-0.033})$ & $14.610 \pm 0.010 \; \
                (^{+0.020}_{-0.030})$ & $12.949 \pm 0.026 \; (^{+0.027}_{-0.026})$ & $13.292 \pm 0.036 \; (^{+0.027}_{-0.036})$ \\
PG0003+158 & $19.369 \pm 0.063 \; \
                (^{+0.068}_{-0.081})$ & $14.540 \pm 0.030 \; \
                (^{+0.110}_{-0.120})$ & $12.627 \pm 0.066 \; (^{+0.133}_{-0.140})$ & $13.170 \pm 0.040 \; (^{+0.090}_{-0.130})$
\enddata
\tablecomments{All the column densities are estimated integrating the \Nav\ profiles over the velocity ranges, [$v_1, v_2$], in Table \ref{tab:vel-range}. \label{tab:column} The first error is the random uncertainty, and the second is the systematic uncertainty associated with isolating Perseus gas from local gas (see Section \ref{subsection:column} for more details).}
\end{deluxetable*}


\subsection{Kinematics of the Disk-Halo Interface Warm-hot Gas}
\label{subsec:Kinematics}

Feedback-driven bulk transportation of matter from the Perseus arm to the disk-halo interface necessarily includes vertical motion. This will include a radial component that will contribute to the observed gas velocities in addition to the effects of galactic rotation and streaming. For example, \citet{Kepner1970} and \citet{Verschuur1973} found that the extraplanar \HI\ emission above Perseus arm splits into inner and outer spiral-like $l$--$v$ branches, which is explained by \citet{Sofu1974} as rising/falling gas motions. Consequently, one might expect absorption from extraplanar Perseus gas to be offset in velocity somewhat from the velocity found in the disk.

In practice, however, the impact of such motions on our observations is hidden by the complexity of the overlapping absorbing components and the coarse spectral resolution, especially, of the COS instrument. We discuss below the apparent optical depth-weighted average Perseus velocities ($v_a$) and the velocity ranges containing 90\% of Perseus absorption ($\Delta v_{90}$). These show no discernible patterns, particularly in the disk-halo interface sightlines, that might be associated with vertically-oriented outflows, but they are limited as diagnostic tools for studying the kinematics of the disk-halo interface gas.

\begin{figure*}
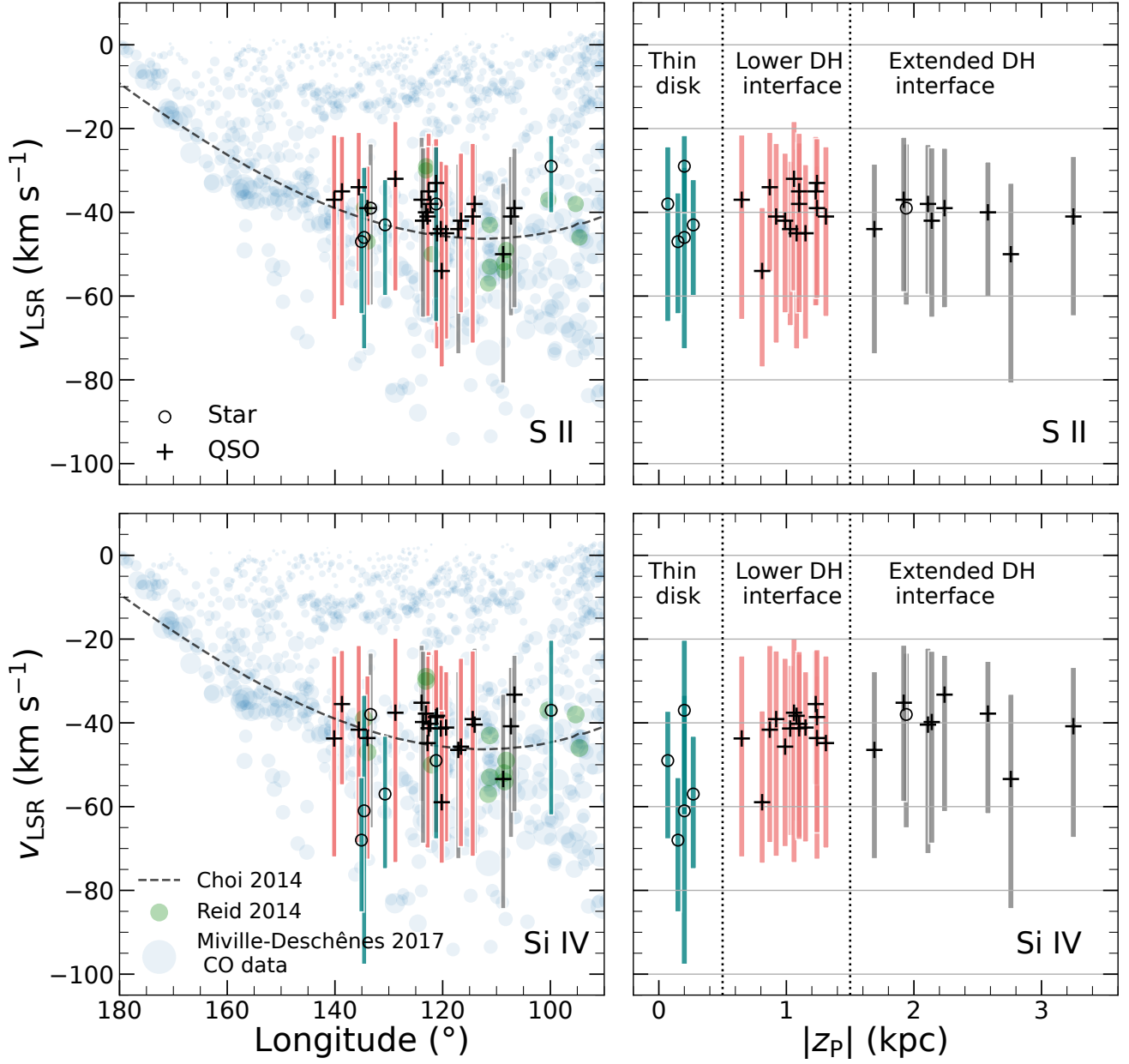

\fig{kinematics.pdf}{0.95\textwidth}{}
\caption{
Kinematics of the \SII\ ({\it top}) and \SiIV\ ({\it bottom}) absorption associated with the Perseus arm against the Galactic longitude $l$ ({\it left}) and absolute mid-arm height \z\ ({\it right}). The average Perseus velocity, $v_a$, and the velocity range containing 90\% Perseus absorption, $\Delta v_{90}$, in \SII\ and \SiIV\ are shown with the plus signs and the vertical bars, respectively. $\Delta v_{90}$ (the vertical bars) are color-coded based on the mid-arm heights, \z, e.g., teal for $z_{\rm P}< 0.5$ kpc, pink for $0.5<z_{\rm P}< 1.5$ kpc, and grey for $z_{\rm P}> 1.5$ kpc. The adopted Perseus velocity track from \citet{Choi2014} is shown as a black-dash line in the left panels, along with CO data from \citet{Miville2017} (in light blue circles) and high-mass star-forming region (HMSFR) data from \citet{Reid2014} (in green circles).
}
\label{fig:kinematics}
\end{figure*}

Figure~\ref{fig:kinematics} shows the $l$ (left) and \z\ (right) distributions of the Perseus velocities in our spectra ($v_a$ and $\Delta v_{90}$) for \SII\ (top panels) and \SiIV\ (bottom panels). We find: 1) \SII\ $v_a$ ranges between $-55$ \kms\ to $-25$ \kms\ with $\Delta v_{90}$ typically spanning over $[-80,-20]$ \kms; 2) except for the sightlines toward HD~14434, HD~13745, and HD~232522, \SiIV\ $v_a$ and $\Delta v_{90}$ are consistent with \SII\ within $\sim 5$ \kms; 3) both ions' average velocities follow the expected $l$--$v$ trend for the Perseus arm material (black-dash line in the left panels of Figure \ref{fig:kinematics}, \citealt{Choi2014}), with a scatter of $\sim 6$ \kms; 4) a Spearman rank correlation test yielded p = 0.7 and 0.2 for \SII\ and \SiIV\ respectively, indicating no statistically significant monotonic correlation between average velocities, $v_a$ and vertical height, \z.

We do note that the average velocity of high ions from Perseus gas in the thin disk sight lines shows significant scatter, and absorption is seen toward more negative velocities. Several of these sightlines, notably toward HD~14434, HD~13745, and HD~232522, may pass through expanding superbubble-like structures \citep[][see Section \ref{subsec:velocity}]{Knauth2003,Cappa2000}. That is in contrast to what is seen toward the disk-halo interface sight lines, where the scatter is relatively smaller and the velocities of absorption do not extend to such high negative velocities. While our ability to trace subtle shifts in the velocity structure in the disk-halo interface is limited by the complexities of the absorption, we clearly do not see the prominent signatures of expansion that are likely seen toward the thin disk stars. Instead, the \SiIV\ and \SII\ profiles are consistent with a cylindrically corotating extension of the Perseus at $|z| > 0.5$ kpc.

\begin{figure*}
    \gridline{\fig{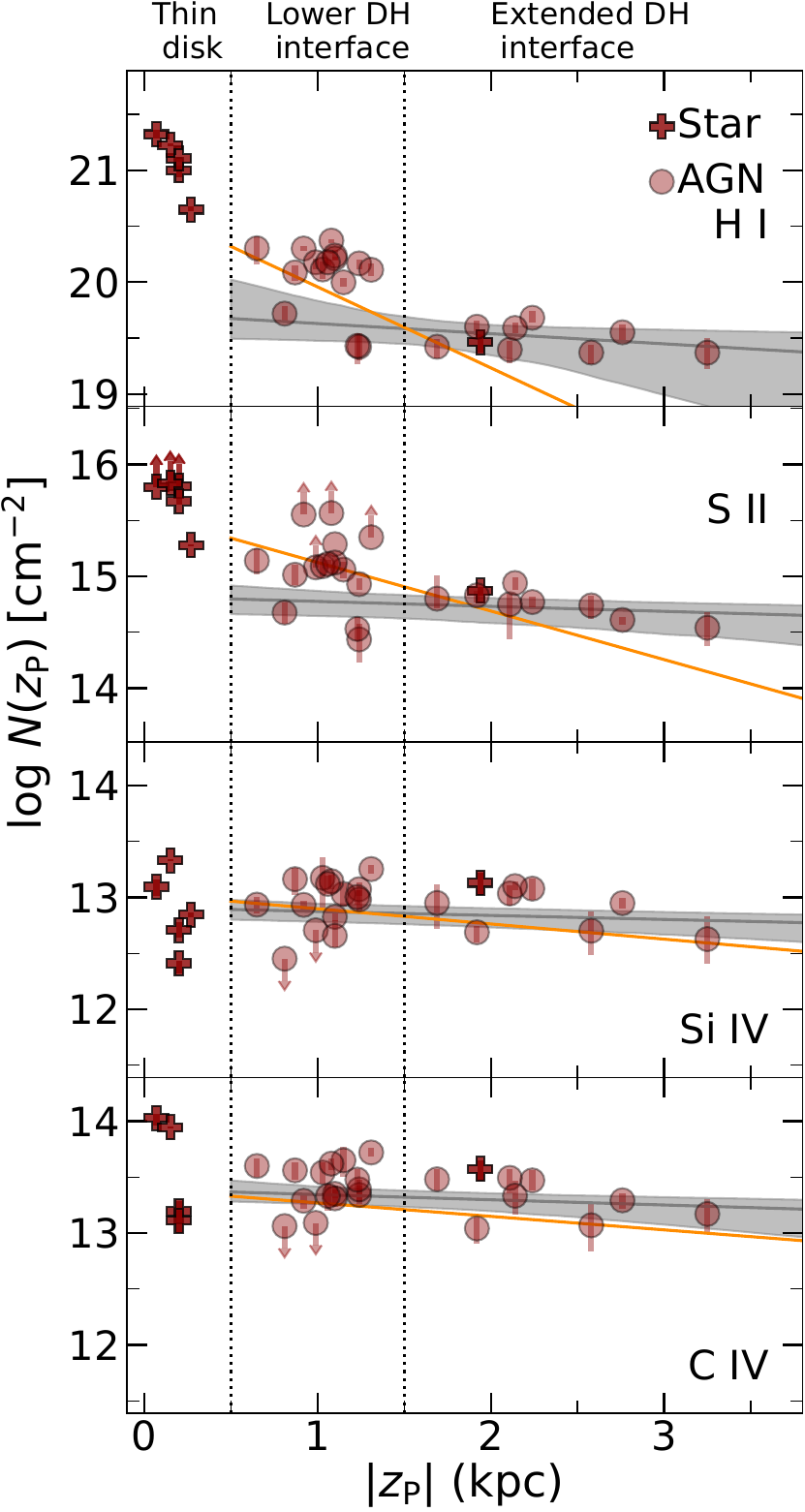}{0.9\columnwidth}{}
              \fig{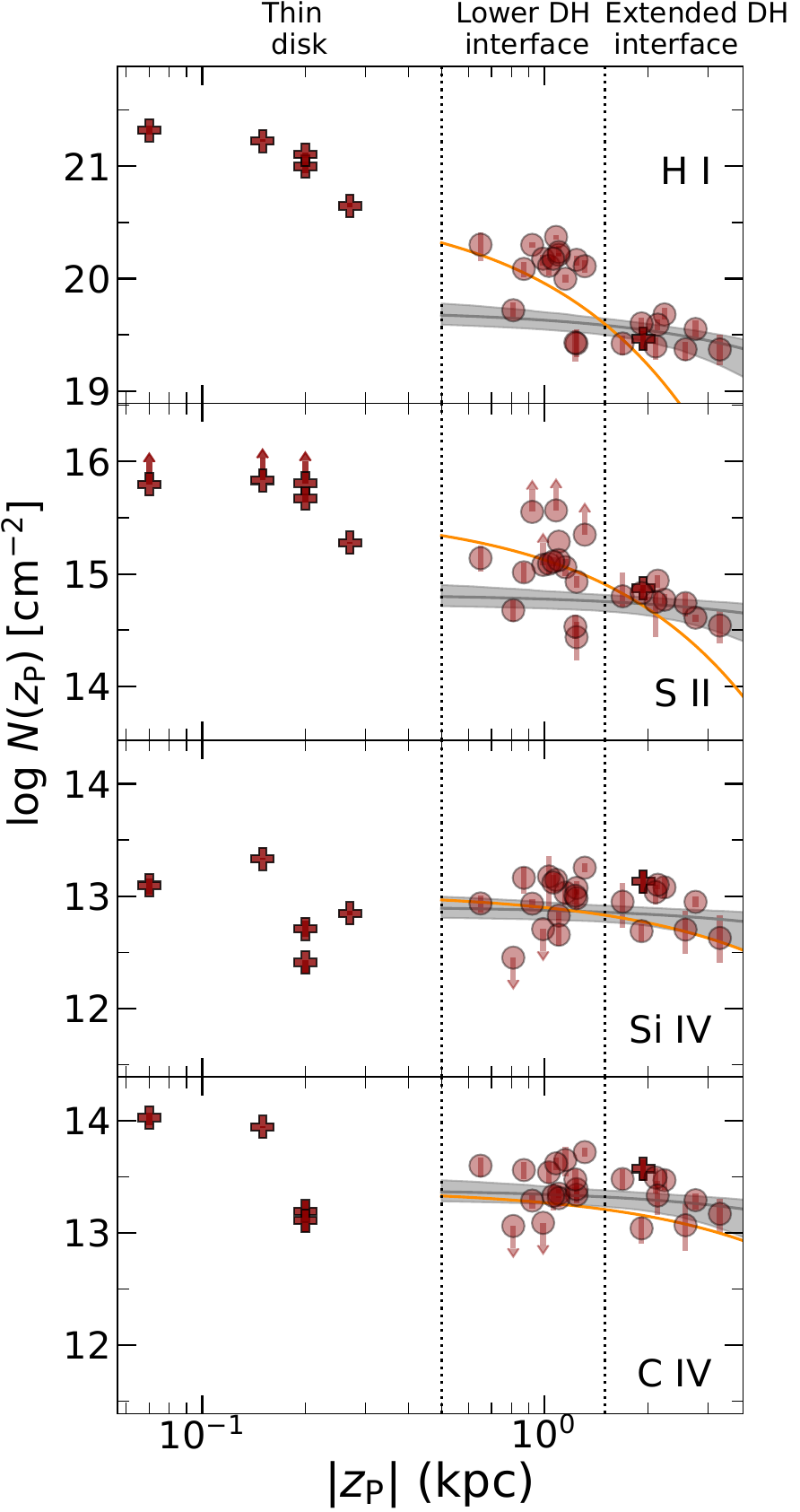}{0.9\columnwidth}{}}
  \caption{Logarithm of \HI, \SII, \SiIV, and \CIV\ column densities, $N(z_{\rm P}$), as a function of \z\ ({\it left}) and $\log$ \z\ ({\it right}). The dark red pluses and circles are estimated $\log N(z_{\rm P})$ in the stellar and AGN sightlines, respectively. The upward and downward arrows indicate lower or upper limits, respectively. The gray solid line in each panel indicates the MCMC best fit to the data at \z$>0.5$ kpc assuming a single exponential model, and the gray shaded region indicates the 95\% confidence interval (CI) of the MCMC fits. The orange solid line indicates the expected column density distribution with exponential scale heights $h_z$ from literature ($h_z = 0.5$ kpc for \HI\ (\citetalias{Dickey1990}); $1.0$ kpc for \halpha\ emission probing similar gas phase as \SII\ (\citealt{Haffner1999}); 3.2 kpc for \CIV, and 3.4 kpc for \SiIV, (\citetalias{Savage2009})).
  \label{fig:colden-zheight}}
\end{figure*}

\begin{deluxetable*}{lCCCCCCCCC}
\tablenum{4} \tabletypesize{\footnotesize}
\tablecaption{MCMC Fit Results of the $N(z_{\rm P})$--\z\ Distributions.\label{tab:MCMC}}
\tablehead{\colhead{Ion} & \colhead{} & \colhead{} &  \colhead{$\log N_0 \,^{\rm a}$}  & \colhead{} & \colhead{} & \colhead{$\sigma_p\,^{\rm b}$}  & \colhead{} & \colhead{} & \colhead{$h_z\,^{\rm c}$}\\ \colhead{ } & \colhead{} & \colhead{} & \colhead{[$\rm{cm}^{-2}$]}  & \colhead{} & \colhead{} & \colhead{}  & \colhead{} & \colhead{} & \colhead{(kpc)}}
\startdata
H I & & & $19.74(^{+0.39}_{-0.13})$ & & & $0.25(^{+0.05}_{-0.04})$ & & & $>1.11$ \\
S II & & & $14.82(^{+0.10}_{-0.07})$ & &  & $0.23(^{+0.06}_{-0.04})$ & &  & $>2.80$ \\
Si IV & &  & $12.92(^{+0.06}_{-0.05})$ & & &  $0.17(^{+0.04}_{-0.03})$ & & & $>4.40$ \\
C IV & & & $13.39(^{+0.09}_{-0.06})$ &  & & 0.$15(^{+0.04}_{-0.03})$ &  &  & $>3.10$\\
\enddata

\tablecomments{For each ion, we fit a single exponential model to the column density data at \z$\geq0.5$ kpc.  (a): best-fit mid-plane column density with 68\% CI, (b): best-fit patchiness parameter with 68\% CI, (c): 5th percentiles of \HI, \SII, \SiIV\ and \CIV\ scale height posterior distributions.}
\end{deluxetable*}

\subsection{Structure of the Extraplanar Warm-Hot Gas}
\label{subsec:column_dist}

We next examine the extraplanar gas distribution, i.e., the \z-distributions of \HI, \SII, \SiIV, and \CIV\ column densities in Figure~\ref{fig:colden-zheight} (plotted vs. linear \z\ in the left panel and log \z\ in the right panel). We remind readers that the sightlines at $|z_{\rm{P}}|<0.5$ kpc probe the thin disk Perseus arm material, whereas at $|z_{\rm{P}}|>0.5$ kpc probe the disk-halo interface below the Perseus arm. We observe distinct trends in the low and high ions, as discussed below.

In \HI\ we observe: 1) $\log N (z_{\rm P})$ sharply drops by $\sim 0.8$ dex up to $|z_{\rm{P}}| \approx 0.5$ kpc; 2) at $0.5<|z_{\rm{P}}|<1.5$ kpc, \HI\ $\log N(z_{\rm P})$ drops by another $\sim 0.5$ dex; 3) at $|z_{\rm{P}}|>1.5$ kpc, \HI\ $N(z_{\rm P})$ does not change at all with increasing \z\ and has a $\sim 0.1$ dex scatter. At $|z_{\rm{P}}|>1.5$ kpc, we find that \HI\ $\log N(z_{\rm P})$ is 0.3--0.5 dex higher than the expectation from a $h_z = 0.5$ kpc exponential scale height (orange solid line in Figure~\ref{fig:colden-zheight}; \citetalias{Dickey1990}). This indicates a possibly larger scale height of the \HI\ in the disk-halo interface below the Perseus arm. Within 0.5--3.3 kpc, \HI\ $\log N (z_{\rm P})$ drops roughly $1$ dex with increasing height---indicating a shallower column density distribution in the disk-halo interface than in the thin disk.

The \SII\ Perseus absorption is saturated in the thin disk. At $0.5<|z_{\rm{P}}|<1.5$ kpc, we see a large scatter ($> 1$ dex) in $N (z_{\rm P})$ with saturated \SII\ in four sightlines near $|z_{\rm{P}}|\sim1$ kpc. To confirm that these saturated absorption features are not associated with the local material, we examine 8 stars in the foreground of the Perseus arm at $D_{\rm{star}}<1.5$ kpc of the Sun. None of these sightlines exhibit Perseus-like \SII\ absorption, which we define as absorption features within velocities $-90<\vlsr<-25$ \kms. This confirms that the saturated absorption arises at distances $D>1.5$ kpc. These saturated \SII\ lines may be associated with dense structures in the disk-halo interface gas below the Perseus arm. At $|z_{\rm{P}}|>1.5$ kpc, \SII\ $N (z_{\rm P})$ has a smaller scatter ($\sim 0.1$ dex) and varies little ($<0.1$ dex) with increasing \z. Again, at $|z_{\rm{P}}|>1.5$ kpc, we measure higher \SII\ column densities than expected from the 1 kpc scale height (the orange solid line in panel 2 Figure~\ref{fig:colden-zheight}) of \halpha\ emission \citep{Haffner1999}, which probes similar gas phase as \SII. Within 0.5--3.3 kpc, the unsaturated \SII\ $N (z_{\rm P})$ drops only roughly $\sim 0.6$ dex with increasing height, which is a shallower decline than \HI.

Unlike the low ions, \CIV\ and \SiIV\ $N (z_{\rm P})$ exhibit much flatter distributions, showing similar column densities and $\sim 0.2$ dex scatter across the entire sampled \z\ range. In the disk-halo interface between $|z_{\rm{P}}|=0.5$--3.3 kpc, there might be a slight decline ($<0.1$ dex) in both high ion column densities with increasing height. Figure~\ref{fig:colden-zheight} shows the slope expected for \SiIV\ and \CIV\ based on the scale heights determined by \citetalias{Savage2009} (3.2 and 3.6 kpc, respectively). We remind again that \citetalias{Savage2009} experiment is different from ours as they measure $N(<z)$.

To test if the $N(z_{\rm P})$--\z\ distributions are exponential in nature, i.e., $N(z_{\rm P}) = N_0 \exp{[-z_{\rm P}/h_z]}$ with a mid-plane density $N_0$, a scale height $h_z$, we model the column densities of each ion using a Markov Chain Monte Carlo (MCMC) fitter. \footnote{Our likelihood function for uncensored data is
$\ln p(N|z_P,N_0,\sigma_p,h_z) = - \frac{1}{2}[(N - N_0 \exp{(\frac{-z_P}{h_z})})\, /\, (\sigma_N^2 + \sigma_p^2)+ \ln {(\sigma_N^2 + \sigma_p^2)}]$, where, $\sigma_N$ is the uncertainty in column density measurements. For upper and lower limits we use the error function $\rm{erf}(w) = (N - N_0 \exp{[-z/h_z]})\, /\, (\sqrt2[\sigma_N + \sigma_p])$ such that $\ln p(N|z_P,N_0,\sigma_p,h_z) = \ln[ 0.5 \pm 0.5\, \rm{erf}(w)]$. We use uniform flat prior on $N_0,\sigma_p,h_z$ for $12<\log N<22, \; 0<\sigma_p<1$, and $0<h_z<10$ kpc.} We also include an ISM patchiness parameter $\sigma_p$ to account for the inherent scatter in the ISM gas distribution \citepalias{Savage2009}. The results are summarized in Table \ref{tab:MCMC}. The $N_0$ and $\sigma_p$ posterior constraints are well-defined, but none of the $h_z$ is constrained. We, therefore, list the $2\sigma$ lower limits (2.5\%) of $h_z$ posterior distributions in Table~\ref{tab:MCMC}, but we caution that these lower limits are prior-dependent. The gray solid line and the gray shaded region in each panel of Figure \ref{fig:colden-zheight} show the best-fit MCMC model and their 95\% CI. In \HI\ and \SII, there is a strong disagreement between our MCMC fits and the flat slab model predictions (orange solid lines). However, despite the poorly constrained scale heights in the high ions, our MCMC fits and the flat slab model predictions do not differ much. This suggests that in the disk-halo interface below the Perseus arm: 1) the neutral/low ionization gas's $N(z_{\rm P})$--\z\ distribution is more complex than a single exponential distribution; 2) the high ions could follow a single exponential distribution with scale height, $h_z\geq3$ kpc.

To summarize, \HI\ and \SII\ have distinct $N(z_{\rm P})$--\z\ distributions compared to the high ions. Both $N_{\rm {H I}}$ and $N_{\rm {S II}}$ sharply decrease with height up to $|z_{\rm{P}}|=1.5$ kpc, and above $|z_{\rm{P}}|>1.5$ the decline is relatively shallow. For the high ions, $N(z_{\rm P})$--\z\ distributions are remarkably flat over the entire \z\ range probed by our sample. The observed $N(z_{\rm P})$--\z\ distributions in the disk-halo interface below the Perseus arm, particularly in the neutral/low ionization gas, are incompatible with the previous exponential scale height characterizations.

One might suspect that the observed shallow $N(z_{\rm P})$--\z\ distributions could result from the uncorrected geometric bias, as the path length through the Perseus gas increases with increasing latitude of our sightlines. However, we show in Appendix \ref{sec:app-C} that the effect of such a geometric bias in $N(z_{\rm P})$ would be minimal. The column densities projected along the Perseus arm's width, $\log[N \cos{|b|}]$, are very similar to the corresponding $N(z_{\rm P})$ values, particularly along the disk-halo interface sightlines. Consequently, the $\log[N \cos{|b|}]$--\z\ distributions are indistinguishable from the $N(z_{\rm P})$--\z\ distributions, and our conclusions remain valid despite the potential geometric effects.


\subsection{Ion ratios}
\label{subsec:Ion-ratio}

Using the Perseus column densities in Table \ref{tab:column}, we estimate \siIVsII\ and \cIVsiIV. The \siIVsII\ values help determine the dominant phase between the warm and warmer/transition temperature gas (see Section \ref{subsection: final}), while \cIVsiIV\ is useful to constrain the ionization mechanisms of the highly ionized gas. The change in the ionization conditions in the extraplanar Perseus arm gas can be assessed from the \z-distributions of these ratios.

\subsubsection{\siIVsII\ vs  \z}
\label{subsubsec:SiIV-SII}

\begin{figure}
    \fig{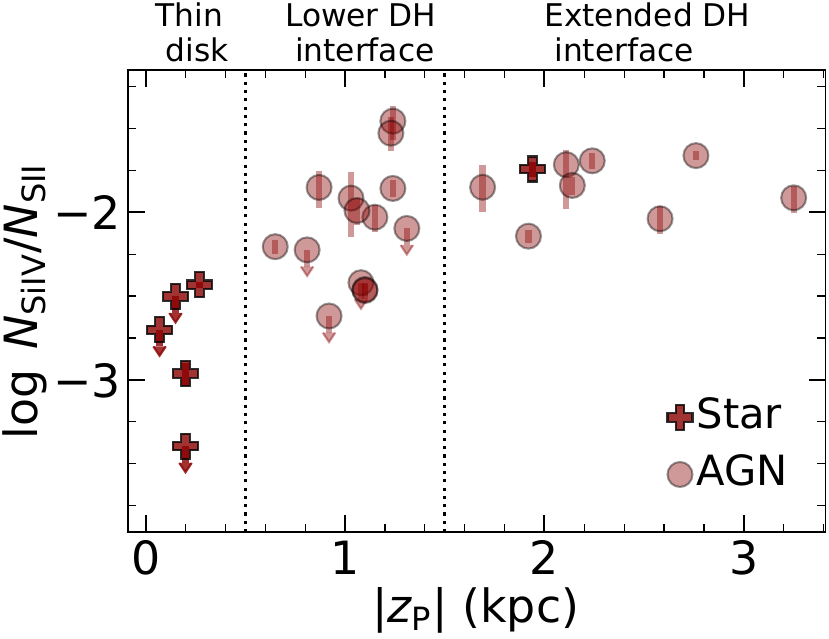}{0.99\columnwidth}{}
    \caption{The \z-distribution of $\log\,$\siIVsII. Symbols are the same as in Figure \ref{fig:colden-zheight}. The data points with downward arrows indicate the upper limits of the ratio due to saturated \SII.
    \label{fig:ratio_SiIV_SII}
    }
\end{figure}

Figure \ref{fig:ratio_SiIV_SII} shows the \z-distribution of  \siIVsII. At $|z_{\rm{P}}|<0.5$ kpc, \SII\ is mostly saturated (3 out of 5 measurements, see panel 2 Figure \ref{fig:colden-zheight}), resulting in \siIVsII\ upper limits at those heights. At $0.5<|z_{\rm{P}}|<1.5$ kpc, \siIVsII\ gradually increases and plateaus at $|z_{\rm{P}}|>1.5$ kpc with \siIVsIIav~$= -1.8 \pm 0.2$. The rise in \siIVsII\ up to 1.5 kpc is primarily due to the large drop in $N_{\text{S II}}$ with increasing \z\ and $N_{\text{Si IV}}$ barely changing with \z\ (see Figure \ref{fig:colden-zheight} panels 2 and 3). At $|z_{\rm{P}}|>1.5$ kpc, \siIVsII\ plateaus owing to the flattened $N_{\text{S II}}$ and $N_{\text{Si IV}}$ distributions at large heights (see section \ref{subsec:column_dist}).

Notably, \siIVsII\ is $\ll1$ at all \z, indicating significantly larger \SII\ column density compared to \SiIV\  as well as \CIV\ (as discussed in section \ref{subsubsec:CIV/SiIV_vs_z}). At $|z_{\rm{P}}|>1.5$ kpc, e.g., in the extended disk-halo interface, the \SII\ column density on average is roughly 60 times higher than \SiIV. Previously, using individual sightlines toward inner and outer Galaxy disk and halo stars at $|z_{\text{star}}|$ ranging between 43 pc to 3 kpc, where path length through the thin disk gas dominates, $\log$ \siIVsII\ is found to be between $-2.6$ to $-1.7$ dex with no obvious dependence on $|z_{\text{star}}|$ (\citealt{Spitzer1992, Spitzer1993, Fitzpatrick1994, Fitzpatrick1997, Howk1999, Brandt1999, Sembach1994, Sembach1995}). Toward globular cluster stars at $|z_{\text{star}}|=5.3$ and 10 kpc, where path length through the disk-halo interface gas dominates, \citet{Howk2012} found $\log$ \siIVsII~$=-1.70 \pm 0.02$, and $-1.48 \pm 0.03$ respectively. These Galactic disk and disk-halo interface \siIVsII\ measurements are fully consistent with our \siIVsII\ values measured below the Perseus arm. Additionally, our and old estimates indicate that \siIVsII\ in the disk-halo interface increases with the increasing height above the plane, but $\ll1$---implying that \SII\ column densities are significantly higher than the high ions. \citet{Howk2012} also showed the disk-halo interface gas in their surveyed directions up to 10 kpc above the Galactic plane is predominantly WNM+WIM---given our observed \siIVsII\ values, this might as well be true for the southern vertical extension of the Perseus arm sampled by our sightlines (also see the discussion in Section \ref{subsection: final}). However, it is important to note that we have no information about a hot phase with $T \gtrsim 10^{6}$ K. X-ray spectroscopy of \textsc{O~vii} absorption has indicated that it can be strong in the thin disk and disk-halo regions of the Milky Way \citep{Yao2005}, and this could be a comparably important phase.

\subsubsection{\cIVsiIV\ vs \z}
\label{subsubsec:CIV/SiIV_vs_z}

\begin{figure}
   \fig{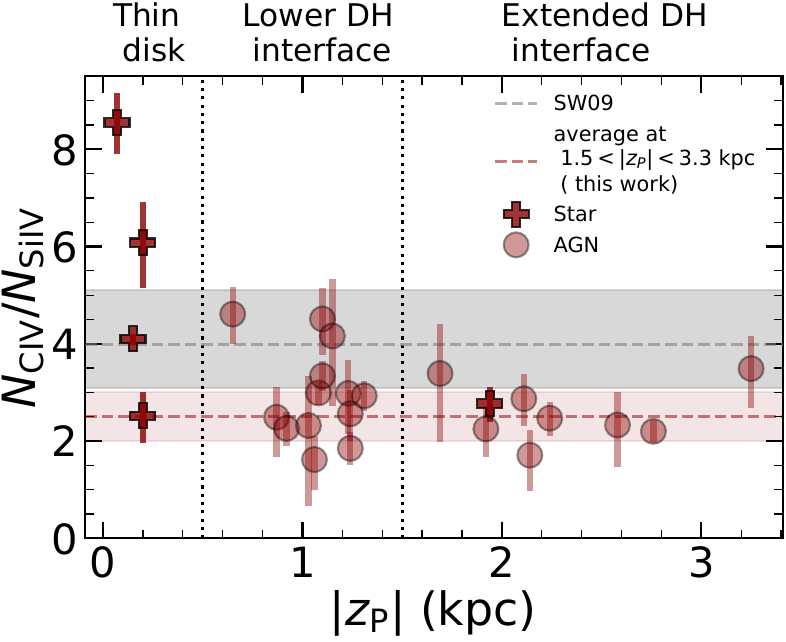}{0.99\columnwidth}{}
  \caption{The \z-distribution of \cIVsiIV. Symbols are the same as Figure \ref{fig:colden-zheight}. The gray-dash line is the Galactic average \cIVsiIVav$=3.98^{+1.15}_{-0.90}$ (the gray shaded region is corresponding $\pm 1\sigma$) from \citetalias{Savage2009}. The red-dash line is \cIVsiIVav$=2.5 \pm 0.5$ (the red shaded region is corresponding $\pm 1\sigma$) at $1.5<|z_{\rm P}|<3.3$ kpc from this work.
  \label{fig:CIVsiIV}}
\end{figure}

Figure \ref{fig:CIVsiIV} illustrates the \z-distribution of \cIVsiIV, revealing a notable pattern: the scatter in \cIVsiIV\ is smaller in the disk-halo interface gas compared to the thin disk. Strikingly, 21 out of 27 \cIVsiIV\ measurements across both thin disk and disk-halo interface sightlines align with the average \cIVsiIVav$=2.5 \pm 0.5$, observed in the extended disk-halo interface at $1.5<|z_{\rm{P}}|<3.3$ kpc.

Six measurements, at $|z_{\rm{P}}|<1.5$ kpc, are more than $1\sigma$ higher than this average. Excluding the outlier toward BD+60~73, which is unusually high, the remaining five measurements elevate the ratio to \cIVsiIVav\ $=3.44\pm1.36$ at $|z_{\rm{P}}|<1.5$ kpc, which is higher than the disk-halo interface average with a larger scatter. This average is, however, consistent with previous works: \cIVsiIVav\ $=3.3 \pm 2.5$ in the thin disk \citep{Lehner2011}, $3.6 \pm 1.3$ \citep{Sembach1992} and $3.98^{+1.15}_{-0.90}$ (\citetalias{Savage2009}) in the disk-halo interface. The large scatter observed in our data at $|z_{\rm{P}}|<1.5$ kpc and in the literature is typical of this ratio. In a study of 40 sightlines in the disk and disk-halo region, \citet{Sembach1997} found a similar mean (median) \cIVsiIV\ $=4.1$ (3.8) with a large rms dispersion ($\sigma = 1.9$). These authors found that the sightlines passing through Radio Loops I--IV, shaped by past supernova explosions, the ratio is significantly higher than the mean/median. Similarly, many sightlines through the Carina star-forming region show \cIVsiIV\ values significantly and consistently smaller than the Galactic average, \cIVsiIVav\ $=3.9$ \citep[see Figure 17 in][]{Tripp2022}. Additionally, component analyses of absorption lines by \citet{Fox2003}, \citet{Tripp2003}, \citet{Lehner2011}, and \citet{Tripp2012} further support this, revealing large scatters ($\sigma=$ 1.5--2.7) in \cIVsiIV\ values within individual components. These observations suggest that the \cIVsiIV\ ratio takes a wide range of values depending on the environment and the physical processes in which \CIV\ and \SiIV\ are created.

While this large scatter in \cIVsiIV\ is typical of ISM measurements and attributable to diverse ionization conditions and mechanisms (see Figure 12 in \citealt{Lehner2011}), our data show a distinct pattern. Both the ratio and the dispersion increase nearer the Galactic plane. The extended disk-halo interface shows an average  $\langle \mbox{\cIVsiIV} \rangle = 2.5$, and a majority of the measurements (21 out of 27) are consistent with this value. However, there is a significant increase in the scatter as one transitions into the lower disk-halo interface and the thin disk. This implies the disk-halo interface sightlines sample gas below the Galactic plane with relatively uniform physical conditions over $0.9<|z_{\rm P}|<3.25$ kpc. The lower-$z$ sightlines sample gas with more diverse conditions.\footnote{ One way in which the conditions in the gas may change is varying degrees of differential depletion of C and Si into dust grains. There is no concrete evidence for dust within the medium hosting the \CIV\ and \SiIV. However, in high depletion environments, a larger fraction of Si could be removed from the gas phase than C \citep{jenkins2009}. Generally, high-latitude sightlines have lower amounts of depletion, low enough that we expect $[{\rm C/Si}] \approx 0$ in the gas phase \citep{savage1996, jenkins2009}. In ``warm disk'' type depletion environments \citep{savage1996} that would be appropriate for most of the sightlines probed here, the differential depletion would yield $[{\rm C/Si}] \approx -0.1$. This is in the opposite sense that we see in the \cIVsiIV\ ratio evolution, where the ratio is higher at low heights. Thus we do not believe dust depletion is driving this variation.}

\section{Discussion}
\label{sec:Discussion}
The main goal of this project is to assess the kinematics, spatial structure, and ionization conditions of the disk-halo interface gas located below the Perseus arm. We detect \SII, \SiIV, and \CIV\ in the UV spectra of 6 OB stars and 23 AGNs. These sightlines pass below the Perseus arm at mid-arm heights between 70 pc and 3.3 kpc. We discuss the implications of our findings below.

\subsection{CGM Contamination to Perseus Absorption is Minimal}
\label{subsec:dis-2}

Every AGN sightline has a significant path length through the Milky Way CGM in addition to the ISM. Consequently, any absorption from the CGM gas along our observed AGN sightlines may overlap with the foreground Perseus arm absorption. Such CGM contamination could lead to inaccurate measurements of Perseus gas column densities. However, we argue that the measured Perseus gas column densities are free from significant CGM contamination based on the following evidence.

First, the estimated Perseus arm column densities toward the AGNs are similar to or smaller than that toward the halo star, PG0122+214 ($|z_{\rm{P}}|=1.9$ kpc). Particularly, in the high ions where $N(z_{\rm P})$--\z\ distributions are very flat over the whole \z-range, the Perseus column densities toward most of the AGNs (15 out of 23) are lower than the halo star sightline by  $\geq1\sigma$ (see Figure \ref{fig:colden-zheight}). The 5 AGN sightlines closest in height (\z = 1.8--2.3 kpc) to the halo star PG0122+214 show high ion column densities 0.03--0.5 dex lower than the stellar sightline, while the \SII\ column densities are $\sim 0.2$ dex lower\footnote{PG0122+214 is a high-velocity runaway main-sequence B star roughly 4--10 kpc (68\% confidence interval from Gaia parallax) below the Galactic plane. As a main sequence B star in a low density region (the halo) and widely separated from its natal environment, we do not expect significant circumstellar material nearby. No excess local emission is detected in either the WHAM \halpha\ or the AKARI FIR maps. The high ion \Nav\ profile toward PG0122+214 is highly consistent with the profiles along two AGN sightlines at different Galactic coordinates, all suggesting the sightline toward this star probes the general ISM and disk-halo absorption like the other sightlines in our sample.}. Along the AGN sightline closest in height to PG0122+214, that toward PG0052+251 ($|z_{\rm{P}}|= 1.9$ kpc), the \SiIV\ and \CIV\ column densities are $0.4$ and $0.6$ dex lower while \SII\ column density is similar to that observed along the halo star sightline PG0122+214. In the presence of substantial low-velocity Milky Way-CGM gas toward these AGN sightlines, one would expect larger column densities toward the AGNs compared to the halo star sightline, which is not the case. The observed differences in column densities toward the AGNs and the halo star are typical of the inherent column density variability in the Milky Way ISM and disk-halo interface at angular separations of 2--25$\degree$ (\citealt{Howk2002}; Tuli et al. 2025, in preparation).

Secondly, we find \siIVsIIav\ $=-1.8\pm0.2$ in the extraplanar gas at $|z_{\rm{P}}|> 1.5$ kpc. \citet{Lehner2020} surveyed the M31 CGM beyond 20 kpc using multiple AGN sightlines and found that \SII\ is not detected, with $\mbox{\siIVsIIav} > -1.09$ in all cases. This is $>4\sigma$ larger than our measurements for extraplanar Perseus gas. While there are only upper limits for \SII, the \SiII\ traces similar gas and were detected more frequently. Lehner et al. (2025) found an average of $\langle \log N_{\rm{Si IV}}/N_{\rm {Si II}} \rangle = -0.10\pm0.35$, $+0.05 \pm 0.37$, and $+0.63\pm0.20$ at $R<50$ kpc, $50<R<150$ kpc and $R>150$ kpc in the M31 CGM, respectively. Similarly, when \SiII\ and \SiIV\ are detected in the COS-Halos survey of low-redshift ($z\sim0.2$) CGM absorption around $L^*$ galaxies, \citet{Werk2013} find $\langle \log N_{\rm{Si IV}}/N_{\rm {Si II}} \rangle = +0.22\pm0.29$. To compare these values to our observations, we must correct for the different intrinsic abundances of Si and S. For extraplanar Perseus gas at $|z_{\rm{P}}|> 1.5$ kpc, we find $ [\SiIV/\SII] \equiv \log (N_{\SiIV}/N_{\SII}) - \log (\rm{Si}/\rm{S})_{\odot} = -2.2$. This result can be directly compared to the CGM $N_{\rm{Si IV}}/N_{\rm {Si II}}$ ratios and is clearly much lower than found in CGM gas. This contrast between the ratios observed along our sightlines and typical CGM values implies that the extraplanar Perseus environment is significantly less ionized than typically observed in circumgalactic gas.

The consistent column densities of the stellar and AGN sightlines (within the ISM scatter) and the dominance of the low ionization phase in the extraplanar gas at $|z_{\rm{P}}|> 1.5$ kpc, therefore, provide compelling evidence that the observed \SII, \SiIV, and \CIV\ absorption originates from the disk-halo interface gas localized below Perseus arm with negligible contribution from the Milky Way CGM toward the AGNs.

\begin{figure}[h]
   \fig{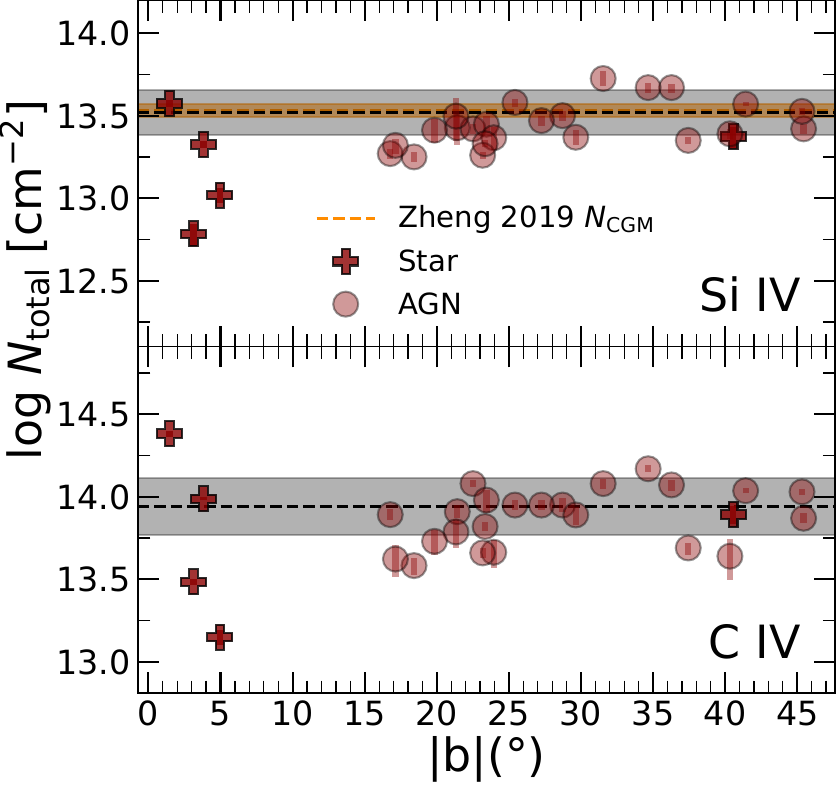}{0.95\columnwidth}{}
  \caption{Total integrated \SiIV\ (top) and \CIV\ (bottom) column density (within $|\vlsr|<100$ \kms) as a function of increasing $|b|$. The grey dash line and the grey shaded region are the average and the standard deviation of $\log N_{\rm total}$ at $|b|>30\degree$, respectively. The orange dash line and the orange shaded box (top panel) are the \citet{Zheng2019} $\log N_{\rm CGM}$ and its 68\% confidence interval.
  \label{fig:col_tot}}
\end{figure}

\subsection{CGM Contamination to the Absorption at $|\vlsr|<100 \ {\rm km\, s^{-1}}$ is Minimal}
\label{subsec:total-column}

The Milky Way-CGM gas has been traditionally indirectly detected as high velocity clouds (HVCs) at $|\vlsr|>100$ \kms\ in AGN spectra \citep{Herenz2013, Richter2017}. However, there is also growing speculation \citep{Zheng2015, Zheng2020} that a substantial portion ($55\%$--$65\%$) of the warm-hot Milky Way CGM gas is concealed within the low velocity gas (i.e., the gas with $|\vlsr|< 100$ \kms) typically attributed to the nearby Milky Way ISM and disk-halo interface gas. \citet{Zheng2019} modeled the all-sky \SiIV\ column densities toward the AGNs with a 1D ($z$-dependent) disk-halo interface density distribution ($ N_{\rm {DH}}$) coupled with a global CGM component ($ N_{\rm{CGM}}$), such that the average $\log N_{\rm {DH}} = 12.11$, and $\log N_{\rm{CGM}} = 13.53$. This suggests that the low velocity \SiIV\ absorption toward the AGNs predominantly originates from the Milky Way CGM gas rather than the disk-halo interface. Subsequent studies using more complex models have argued that the contributions to the \SiIV\ and \CIV\ column densities from the disk-halo interface and the CGM gas are similar, $N_{\rm {DH}} \approx N_{\rm {CGM}}$ \citep{Qu2019, Qu2020, Qu2022}. These works collectively argue that theISM and disk-halo interface  high ion absorption observed along the AGN sightlines include significant contributions from the Milky Way's CGM.

However, the QuaStar survey \citep{Bish2021} measured the column densities in the direction of 30 halo star-AGN pairs with angular separations $<2\degree$, finding minimal excess \CIV\ absorption toward the AGNs compared to the foreground halo stars. They argued that the Milky Way CGM lacks warm gas in general, which is in contradiction with \citet{Zheng2019}, \citet{Qu2019} and \citet{Qu2020, Qu2022}. However, \cite{Qu2022} argued that a radially-extended disk, instead of a vertically-extended one, might result in low \CIV\ CGM column densities toward the QuaStar AGNs.

We find no observable CGM absorption toward the AGNs at the Perseus velocities within $-90<\vlsr<-25$ (see Section \ref{subsec:dis-2}). In principle the low-velocity absorption within $-25<\vlsr<+50$ \kms\ in these directions may still include CGM contributions. To investigate this, we estimate the \SII, \SiIV, and \CIV\ integrated total column densities, $N_{\rm{total}}$, within $|\vlsr|< 100$ \kms, encompassing both Perseus and other velocity components. When examining the full velocity range, we finally note that there is minimal absorption at $\vlsr>50$ \kms\ toward our AGN sightlines, suggesting that the high positive velocity components make negligible contributions to the observed $N_{\rm{total}}$ differences.

Due to \SII\ saturation, we focus only on the high ion $N_{\rm{total}}$ vs. $|b|$ distribution of Figure \ref{fig:col_tot}. We find: 1) at $|b|>10\degree$, \CIV\ and \SiIV\ total column densities exhibit $\sigma \sim0.16$ and 0.13 dex scatter, respectively; 2) while the \CIV\ $N_{\rm{total}}$ shows no significant dependence on $|b|$ (Pearson correlation coefficient, $r = 0.34$, p-value $= 0.1$), the \SiIV\ $N_{\rm{total}}$ exhibits a marginal correlation with $|b|$ ($r = 0.4$, p-value $= 0.05$); 3) consistent with the trend seen in Perseus gas (Figure \ref{fig:colden-zheight}), $N_{\rm{total}}$ along the AGNs are consistent with the halo star sightline within the ISM scatter. There are 5 AGN of the 8 at $|b|>30\degree$ toward which we observe \SiIV\ column densities higher than that toward the halo star value by up to 0.35 dex (smaller in excesses in \CIV). These are all in a similar region of the sky, projected within $\approx15^\circ$ of one another and may trace a small excess in the local \SiIV\ gas column (though not much in \CIV). However, their apparent excess is also consistent with the mean and scatter observed in the $N_{\rm{total}}$ distribution of all sightlines at $|b|>10\degree$. Therefore, any difference in $N_{\rm{total}}$ between the halo star and the AGN sightlines is primarily due to the ISM scatter. That is, the AGN sightlines show no evidence for a CGM contribution to the column densities.

\begin{figure}[h]
   \fig{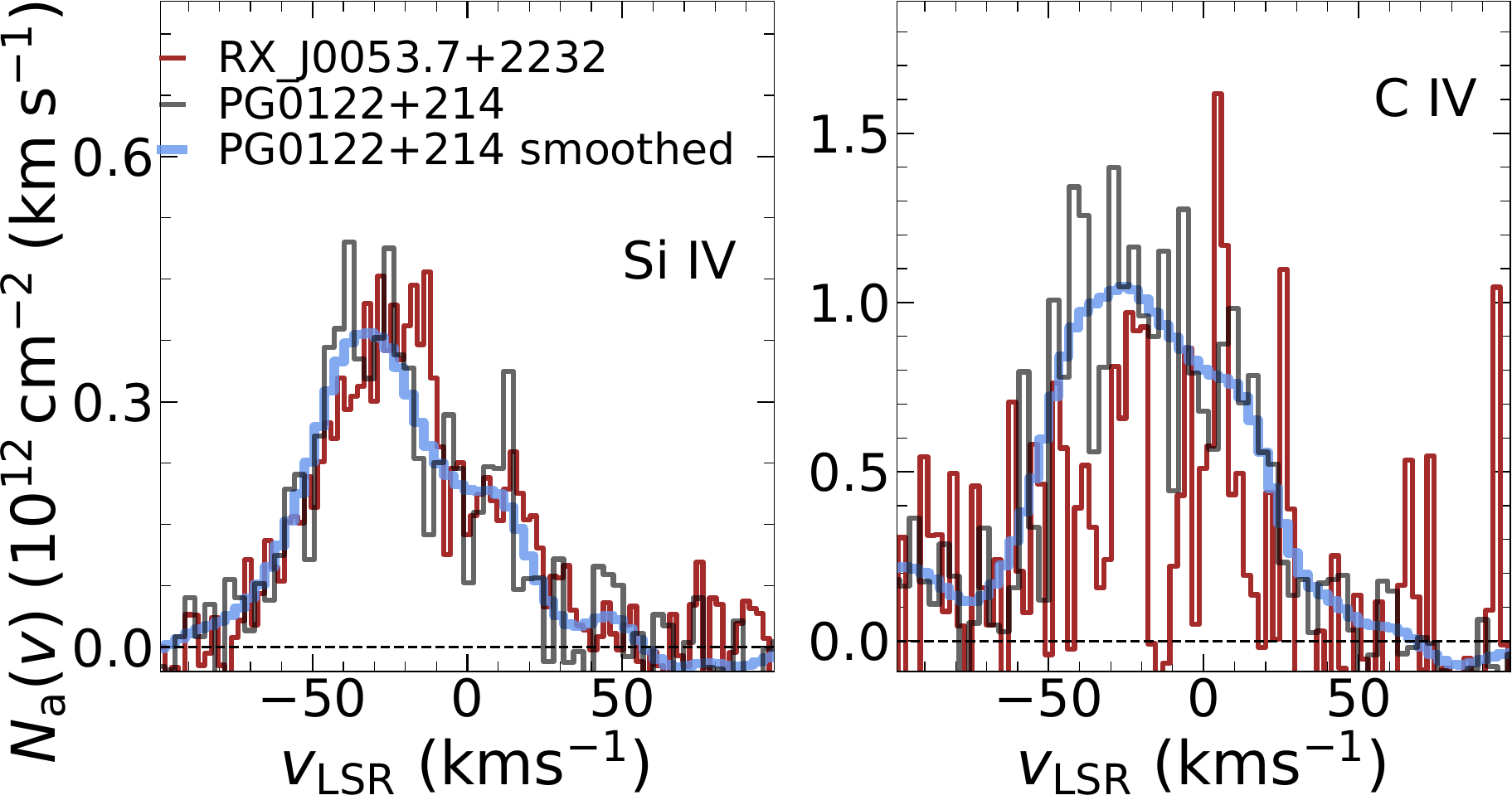}{1\columnwidth}{}
  \caption{Comparison of the \SiIV\ and \CIV\ \Nav\ profiles toward the AGN RXJ0053.7+2232 and halo star PG0122+214. These two sightlines pass below the local and Perseus arm material at different longitudes ($123.64\degree$ and $ 133.37\degree$, respectively) but at similar latitudes ($-40.33 \degree$ and $ -40.57\degree$, respectively). We also show in blue the PG0122+214 $N_a(v)$ profile based on STIS E140M observations that have been smoothed to the equivalent of the COS resolution. The smoothing assumes a pure Gaussian kernel. This allows a direct comparison between the stellar (smoothed STIS E140M) and AGN COS spectra.
  \label{fig:pgvs2232}}
\end{figure}

\begin{figure}[h]
   \fig{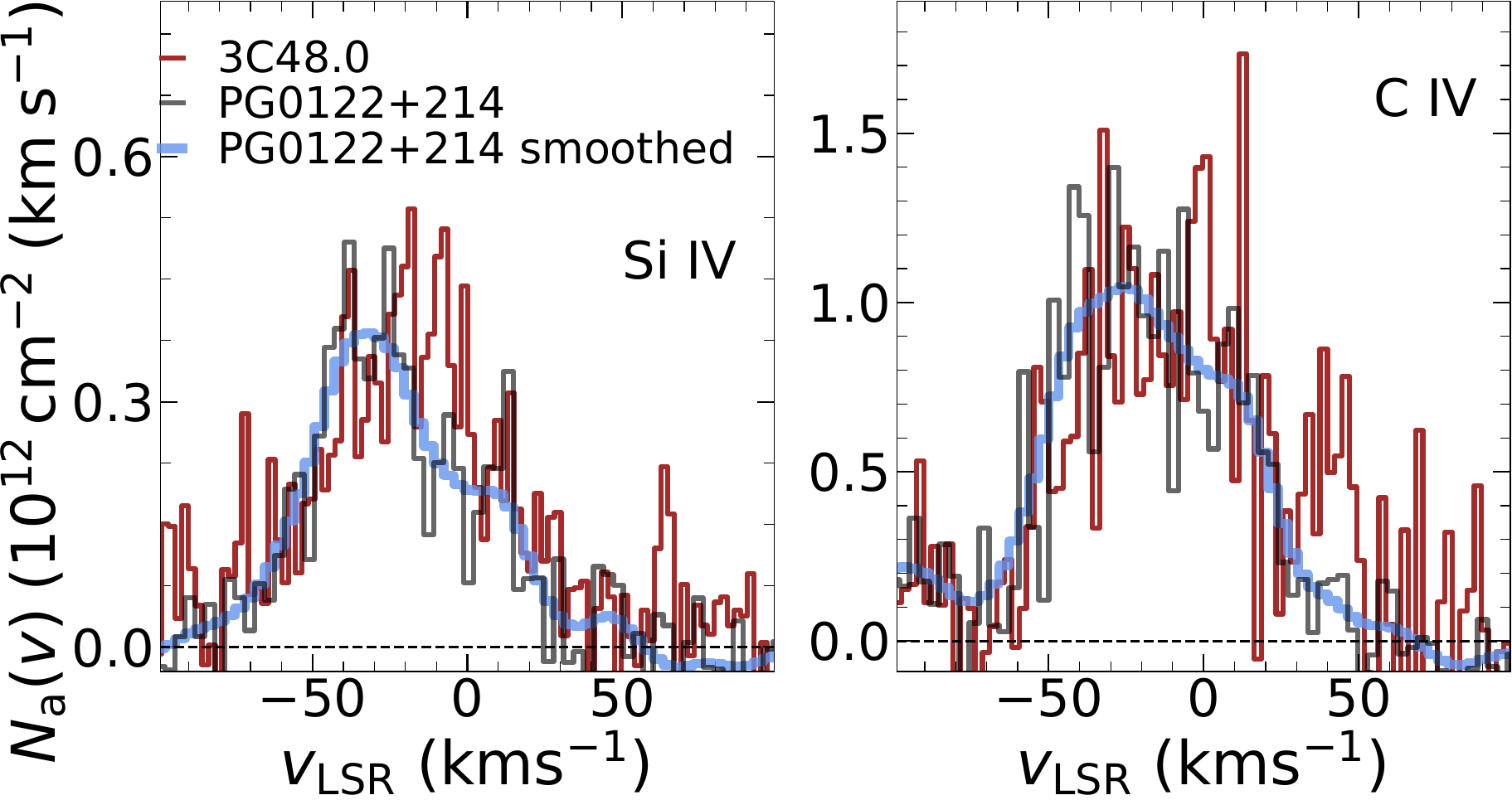}{1\columnwidth}{}
  \caption{Comparison of the \SiIV\ and \CIV\ \Nav\ profiles toward the AGN 3C48.0 and halo star PG0122+214. We show the PG0122+214 STIS E140M $N_a(v)$ profile in grey and in blue we show the STIS $N_a(v)$ profile smoothed to COS resolution). These two sightlines pass below the local and Perseus arm material at similar longitudes ($133.96\degree$ and $ 133.37\degree$, respectively) but at different latitudes ($-28.72 \degree$ and $ -40.57\degree$, respectively).
  \label{fig:pgvs48}}
\end{figure}

For example, the AGN sightline RXJ0053.7+2232 [$(l,b) = (123.64 \degree, -40.33 \degree)$] exhibits similar total high ion column density (within the ISM scatter) as the halo star sightline toward PG0122+214 [$(l,b) = (133.37\degree, -40.57\degree)$]. We compare the \Nav\ profiles for \SiIV\ and \CIV\ toward these sightlines, which have similar latitudes, in Figure \ref{fig:pgvs2232}. This figure shows both the STIS-observed \Nav\ profile toward PG0122+214 and a version that has been smoothed by a Gaussian kernel to match the resolution of the COS AGN observations. Similarly, Figure \ref{fig:pgvs48} shows a comparison of the AGN sightline toward 3C48.0 [$(l,b) = (133.96\degree, -28.72\degree)$] and PG0122+214 [$(l,b) = (133.37\degree, -40.57\degree)$], which are at very similar longitudes but different latitudes. These two sightlines exhibit similar total high ion column densities and show no difference in their high ion \Nav\ profiles, particularly at low velocities.


\begin{figure}[h]
   \fig{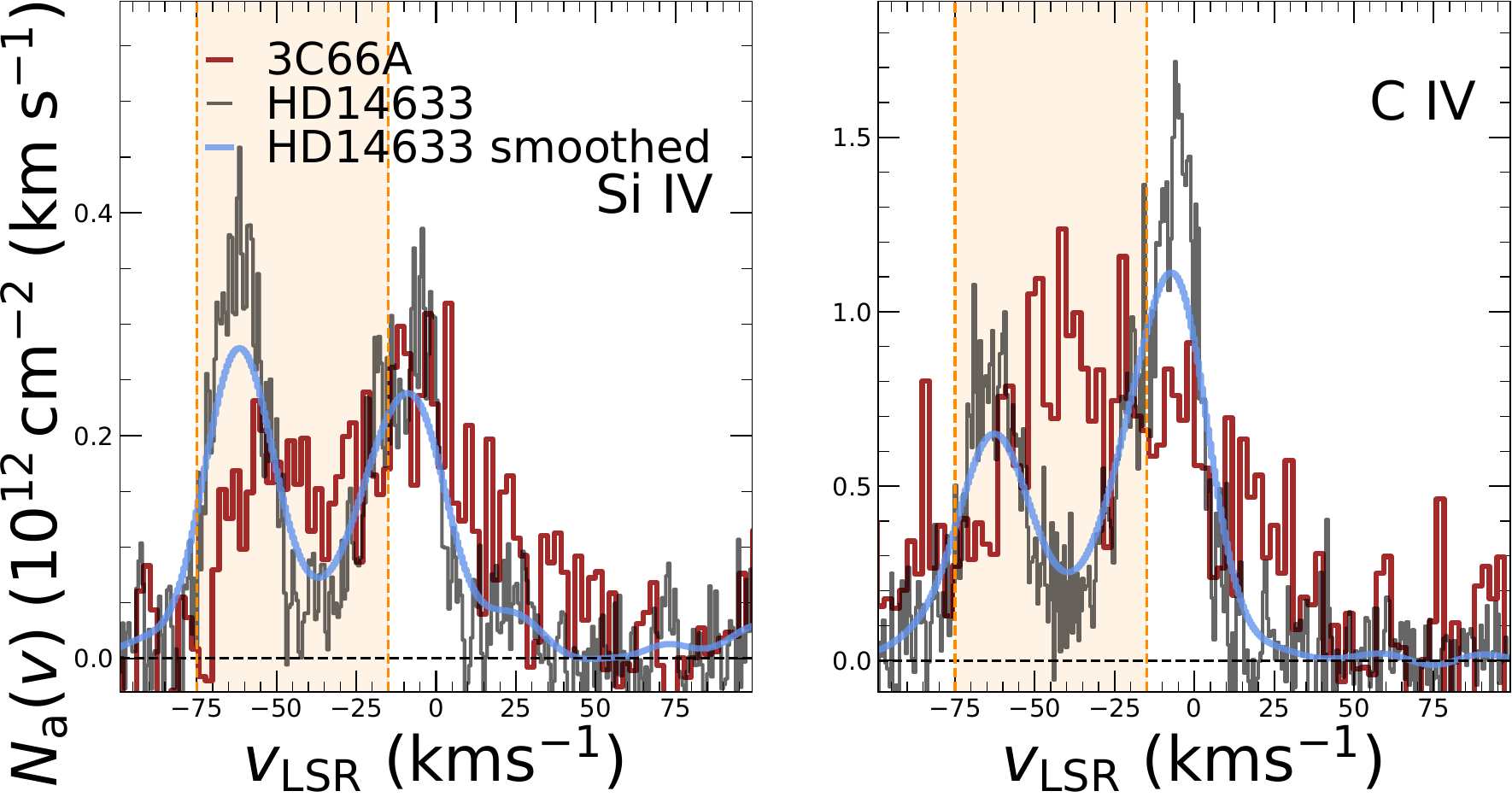}{1\columnwidth}{}
  \caption{A comparison of the \SiIV\ (left) and \CIV\ (right) \Nav\ profiles toward the AGN 3C66A and the nearby star HD~14633, separated by $1.55\degree$ on the sky. We show the HD~14633 STIS E140H $N_a(v)$ profile in grey and in blue we show the STIS $N_a(v)$ profile smoothed to COS resolution. The orange band denotes the velocity range for Perseus absorption in these directions. Unlike the AGN, the star lies in the inter-arm region between the local gas and the the near side of the Perseus arm. Thus, it lacks some of the intermediate velocity absorption from most of the Perseus arm, resulting in the observed differences in the \Nav\ profiles over $-75$ to $-15$ \kms\ (orange shaded region).
 \label{fig:HDvs66A}}
\end{figure}

Additionally, we show in Figure \ref{fig:HDvs66A} a comparison of the \Nav\ profiles for \CIV\ and \SiIV\ toward the AGN sightline toward 3C66A [$(l,b) = (140.14\degree, -16.77\degree)$] and those seen toward the nearby star HD~14633 [$(l,b) = (140.78\degree$, $-18.2\degree)$ and $D_{\rm {star}} = 1.5$ kpc].
This star is not in our sample, as its distances places it at least partially in front of the Perseus arm. HD~14633 and 3C66A are  separated on the sky by only $1.55\degree$ and have total high ion column densities consistent within $\sim0.13 $ dex (typical of ISM scatter at $\sim2$ deg angular separation, Tuli et al. 2025, in preparation). However, though the total columns are similar, the high ion \Nav\ profiles along these two sightlines show significant differences over $-75$ to $+50$ \kms. The star is in the inter-arm region, close to the near side of the Perseus arm, whereas the AGN sightline passes over the whole Perseus arm. We argue that the differences in these sightlines' \Nav\ profiles are due to different path lengths through the Perseus material and not because of the large CGM path length of the AGN sightline, which would result in much larger column density differences than observed between these two sightlines.

These three examples demonstrate that there is significant variability in the low-velocity ISM absorption toward stellar / AGN sightlines relatively closely spaced on the sky (Figures \ref{fig:pgvs48}, \ref{fig:pgvs2232}, and \ref{fig:HDvs66A}). This strongly suggests that the column density differences between the AGN and halo stellar sightlines likely originate from the intrinsic variability in the ISM and disk-halo interface gas distribution rather than from CGM gas beyond the stellar distance.

\begin{figure}[h]
   \fig{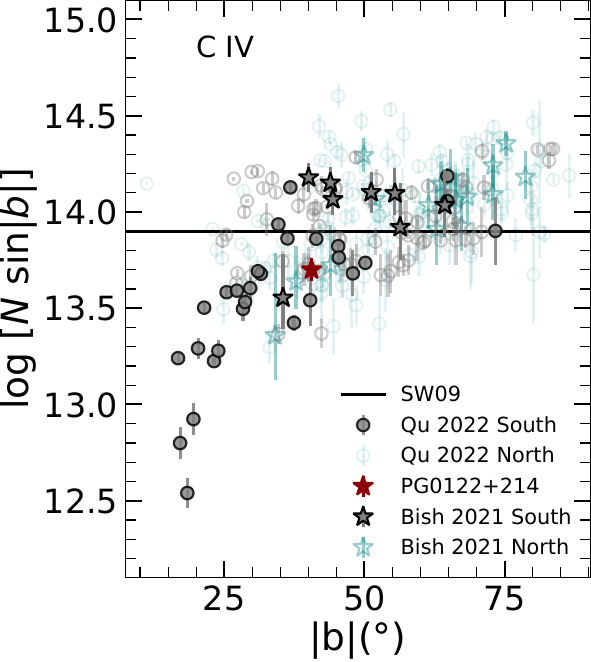}{.95\columnwidth}{}
  \caption{$\log\,N \sin |b|$ vs $|b|$ distribution of \CIV\ toward the southern AGNs in \citet{Qu2022} shown as the gray circles (filled ones are the southern AGN sightlines within $95\degree<l<145\degree$ and open ones are in other directions), southern halo stars (gray stars) in \citet{Bish2021} and our halo star PG0122+214 (red star). The teal open circles and stars in the background are $\log\,N \sin |b|$ measured toward the Northern AGNs in \citet{Qu2022} and halo stars in \citet{Bish2021} respectively. $\log\,N \sin |b|$ measured toward halo stars are consistent with that toward AGNs within the ISM scatter.
  \label{fig:total_col_south}}
\end{figure}

To further test for any statistically significant differences in column densities between halo stars and AGNs, we compare the \CIV\ $\log\, N \sin |b|$ vs $|b|$ distribution from southern halo stars \citep{Bish2021} to our halo star PG0122+214 and the southern AGN sample from \citet{Qu2022} in Figure \ref{fig:total_col_south}.\footnote{ We note that \citet{Bish2021} reported integrated column densities within $|v_{\rm LSR}| < 150$ \kms. After carefully inspecting their \Nav\ profiles, we find minimal absorption beyond $|v_{\rm LSR}| > 100$ \kms, confirming that the absorption along their halo star sightlines arises primarily within $|v_{\rm LSR}| < 100$ \kms. Hence, the $N \sin |b|$ values from the \citet{Bish2021} sample can be directly compared to those from the \citet{Qu2022} AGN sample and our halo star.} Visually, the column densities toward halo stars exhibit similar behavior to the AGN sample. Most notably, column densities toward southern AGNs ($-46\degree<b<-25\degree$) within $95\degree<l<145\degree$ (filled gray circles in Figure \ref{fig:total_col_south}) align closely with those toward halo stars. The $N\sin|b|$ distribution shows comparable $|b|$ dependence for both southern halo stars and southern AGN sightlines, with linear regression slopes of 0.31 and 0.37, respectively (see Appendix \ref{sec:app-D}). A t-test of these slopes indicates no statistically significant distinction in the $N\sin|b|$ vs. $|b|$ relation between southern AGN and southern halo star sightlines ($p=0.09$).

An analysis of the residuals (observed $N\sin|b|$ - the linear model) presented in Appendix \ref{sec:app-D}, reveals no dependence on $d_{\rm star}$ or $|b|$ for either stellar or AGN sightlines (regression slope = 0, p-value = 1). While $N\sin|b|$ increases more steeply with $|b|$ in northern halo star sightlines compared to northern AGN sightlines, residual analysis again shows no $d_{\rm star}$ or $|b|$ dependence in either stellar or AGN sightlines. Moreover, the distribution of the residuals with $|b|$ in halo star sightlines is consistent with AGN sightlines in both northern and southern regions. Two-sample KS tests found no statistically significant differences in column density distributions between halo stars and AGN sightlines in either hemisphere (North: $p = 0.7$; South: $p = 0.1$; both $p > 0.05$). These findings collectively suggest that $N\sin|b|$ values in halo stars and AGN sightlines are consistent within ISM scatter, implying minimal CGM absorption along AGN sightlines. Therefore, the observed column density differences between AGN and halo star sightlines in our sample primarily reflect variations in foreground ISM and disk-halo interface gas.

\subsection{A Complex Spatial Structure with Distinct Physical Conditions below the Perseus Arm}
\label{subsection: final}

In the absence of the Milky Way CGM's contribution in the gas with $|\vlsr|<100$ \kms, we can now discuss our results solely in the context of the disk-halo interface gas in the vertical extension of the Perseus arm.

As mentioned previously, using the diffuse emission from the warm neutral and ionized medium, as well as absorption spectra of hundreds of halo star sightlines (and a handful of AGN sightlines) throughout the Galaxy, early surveys predicted simple exponential gas distribution in the disk-halo interface, i.e., the flat slab model, with temperature-dependent scale heights of the different gas phases (\citealt{Lockman1986, Dickey1990, Haffner1999, Bowen2008, Savage2009}).  However, more recent works using hundreds of AGN sightlines reported that the disk-halo interface absorption toward the AGNs, particularly in the transition temperature ions, might suffer contamination from the low velocity ($|\vlsr|<100$ \kms) CGM gas, therefore requiring more complex models that include a global CGM component as well as the exponential disk-halo interface model to explain the observed column densities toward the AGNs (\citealt{Zheng2019, Qu2019, Qu2020, Qu2022}). Our experiment avoids the ambiguities created by using total integrated sightline column densities, $N(<z)$, by assessing the column density distribution, $N(z_{\rm P})$, directly below the Perseus arm at heights between \z\ $= 70$ pc and 3.3 kpc using 6 stellar and 23 AGN sightlines. Our experiment sheds light on the vertical structure, kinematics, and ionization condition of the disk-halo interface gas below the Perseus arm.

We detect a significant amount of \HI, \SII, \SiIV, and \CIV\ up to 3.3 kpc below the Perseus arm. We find that the vertical structure of the Perseus arm is complex, particularly for the neutral/low ionization gas. \HI\ and \SII\ column densities decline sharply up to \z\ $= 1.5$ kpc, and over \z\ $=1.5$--3.3 kpc, their $N(z_{\rm P})$--\z\ distributions are relatively shallow---significantly deviating from the flat slab fits of the earlier works (see Figure \ref{fig:colden-zheight}). Additionally, \HI\ and \SII\ $N(z_{\rm P})$--\z\ distributions are significantly different from the high ionization gas, also reported by \citetalias{Savage2009}. Like \citetalias{Savage2009}, we find the high ion $N(z_{\rm P})$--\z\ distributions are very flat for the entire range of \z\ probed by our sample (see \ref{fig:colden-zheight}). A single exponential scale height model provides a poor fit for \HI\ and \SII, whereas the highly ionized gas distribution may be exponential, but a scale height $h_z>3$ kpc is required (see Table \ref{tab:MCMC}).

To gain more insight into disk-halo interface gas below the Perseus arm, we investigate the gas kinematics. We find that the high and low ionization gas kinematics resemble the Galactic rotation, roughly spanning over $-20$ to $-75$ \kms. Any vertical fountain flow of roughly few tens of \kms\ \citep{Kim2017} have a small (10--20 \kms) projected line of sight velocity along the disk-halo interface sightlines, which is possibly offset by the projected halo rotation lagging behind the disk rotation. The lack of vertical motion in the gas may also imply that any past fountain flow has significantly slowed down over time, and feedback in this arm region may be currently quiescent. Therefore, our data, particularly in the disk-halo interface sightlines, are unable to capture any vertical motion or rotation lag in the gas below the Perseus arm.

Another key finding is the changing physical and ionization conditions of the Perseus arm gas from the thin disk to the disk-halo interface. We find: 1) in the thin disk, the low and high ionization gas kinematics are very different; however, they are very similar in the disk-halo interface; 2) \siIVsII\ values increase from the thin disk to the disk-halo interface and do not vary at all with height in the disk-halo interface; 3) \cIVsiIV\ values also do not vary at all with height in the disk-halo interface, and the observed scatter in this ratio significantly decreases from the thin disk to the disk-halo interface. These observations indicate that the processes governing the low and high ionization gas phases are very different in the thin disk than in the disk-halo interface.

Secondly, we measure \siIVsIIav$=-1.8$ in the disk-halo interface gas below the Perseus arm, implying the \SII\ column density on average is $\sim 60$ times higher than \SiIV---this is very different from the typical ionization state of the CGM gas observed at low impact parameter ($<15$ kpc) in the M31 and other galaxies \citep{Lehner2020, Werk2013}. \SII\ is primarily produced in the warm neutral medium (WNM) (with ionization fraction, $\chi_{\text{S II}}=1$) and the warm ionized medium (WIM) ($\chi_{\text{S II}}= 0.8$) \citep{Howk2006, Howk2012}. In contrast, \SiIV\ predominantly resides in the more ionized and hotter, $10^{4.2}$--$10^{5.1}$ K, gas phase with $\chi_{\text{Si IV}} \leq 0.38$ under collisional ionization equilibrium (CIE) conditions ($\chi_{\text{Si IV}} \leq 0.06$ for non-equilibrium collisional ionization, NECI) \citep{Gnat2007}. Using the average column density, $N(\rm X)$, in the extended disk-halo interface (\z\ $>1.5$ kpc), ion fraction, $\chi(\rm X)$, and assuming solar abundance, $A(\rm X)$ \citep{Asplund2009}, we can estimate the associated hydrogen column density of a gas phase traced by the ion, X, e,g. $N(\rm H)_{\rm X} = N(\rm X) A(X)^{-1} \chi(X)^{-1}$. We find: 1) $\log N(\text{H})_\text{S II}$ = 19.7, which represents $N(\rm H)$ in the WNM+WIM gas; 2) $N(\text{H})_\text{Si IV}>17.8$ for CIE, $>18.6$ for NECI. The average $\log N(\text{HI})= 19.5$ at $|z_{\rm P}| > 1.5$ kpc in this work represents the WNM. Therefore, in the extended disk-halo interface below the Perseus arm, the $N(\text{H})$ associated with the WIM is $N$(WNM+WIM)$- N$(WNM) = 19.3, implying that the WNM column density is higher than the WIM. However, $\log N(\text{H})_\text{Si IV}$ is unconstrained in this work and a hot (\textsc{O~vii}-bearing) phase could be comparably or even more dominant which is also unconstrained in this dataset.

Thirdly, we find that \cIVsiIVav$=2.5 \pm 0.5$ in the disk-halo interface gas at \z$>1.5$ kpc. The two aspects of this observed ratio are: a) it is below the Galactic average of 3.98; and b) it has a significantly smaller scatter (0.5) than typically observed ($\sim2$, e.g., \citealt{Fox2003, Tripp2003,Lehner2011,Tripp2012})---indicating that our sightlines are probing similar physical conditions in the extraplanar gas spanning over $0.9<|z_{\rm P}|<3.25$ kpc (see Figure \ref{fig:CIVsiIV}). Theoretically, several models depending on the gas temperature can produce the observed low \cIVsiIV\ in the disk-halo interface below the Perseus arm (see Figure 12 \citealt{Lehner2011}). An NCIE \citep{Gnat2007} or shock ionization (SI) model \citep{Gnat2009} can reproduce the observed low \cIVsiIV\ values at gas temperature $T = 2 \times 10^4$ K. Alternatively, a CIE model \citep{Gnat2007} with gas temperature $T = 7 \times 10^4$ K or a turbulent mixing layer (TML) model with 25 \kms\ velocity shear \citep{Slavin1993} at gas temperature $T = 1.6 \times 10^5$ K can also produce the observed low \cIVsiIV\ values. Similarly, cooling flows from the hot ($>10^6$ K) phase to the cool phase ($10^4$ K) can produce the \CIV/\SiIV\ values observed in the disk-halo interface below the Perseus arm in this work, however we also need to measure the other high ion ratios e.g., \CIV/\OVI, and \OVI/\NV\ to constrain these models \citep{Wakker2012, Tripp2022}. Additionally, the observed kinematic similarities between the low and high ionization gas suggest that \CIV\ and \SiIV\ is residing in the mixing layers or interfaces between the hot ($>10^6$ K) and \SII\ bearing cool ($10^4$ K) gas phases; however the existing interface or mixing layer models typically predict a higher \CIV/\SiIV\ ratio than we have observed (see Figure 18 \citealt{Tripp2022}). While all of these aforementioned processes can be traced back to the interaction of the energetic hot feedback material with its surroundings, it is evident from the Figure 12 of \citet{Lehner2011} that the observed disk-halo interface \cIVsiIV\ values are incompatible with the models of photoionizing radiation from the hot stars or the cooling hot plasma.

With minimal contribution from the low-velocity Milky Way-CGM gas along our AGN sightlines, the observed spatial structure and ionization conditions in the disk-halo interface below the Perseus arm may be a consequence of stellar feedback from the arm or material accretion into the arm from the CGM. Our survey excludes regions well below the arm where IVCs were detected in \HI\ 21 cm emission (see Figure 12 of \citealt{Wakker2001}); however, a detailed metallicity assessment in this region could further provide useful constraints. Our sightlines pass through regions of multiple supernova remnants, super-bubbles, and supershells located in and below the arm \citep{Thornton1998, Saud2014}. These structures can influence the gas distribution below the Perseus arm by propelling a substantial amount of arm material to considerable heights. Such phenomena are characteristic of all spiral arms, where Galactic fountains/chimneys occur frequently \citep{Bregman1980,Norman1989, Kim2017, Kim2018}---consequently significantly altering the disk-halo interface's gas distribution and ionization conditions locally above/below the arms.

\section{Summary}
\label{sec:Summary}
We use HST STIS and COS UV absorption spectroscopy of 6 stars and 23 AGNs projected behind the Perseus arm within $95 \degree < l <145 \degree$, and $|b|<46\degree$ to detect \SII, \SiIV, \CIV\ and the associated \HI\ 21 cm emission in the disk-halo interface below the Perseus arm. Our key findings are:
\begin{enumerate}

\item  We have detected warm-hot disk-halo interface material from \z\ $=70$ pc to 3.3 kpc below the Perseus arm. Our AGN sightlines show negligible contribution from the CGM gas, implying that all of the absorption toward the AGNs is arising from the thin disk and disk-halo interface.

\item The neutral and low ionization gas column density distribution below the Perseus arm is complex. \HI\ and \SII\ column densities sharply drop with \z\ up to 1.5 kpc and do not vary much after. In contrast, the high ion column densities do not vary significantly over the whole \z\ range probed by our sample. Our data show that \HI\ and \SII\ column densities do not conform to single exponential distributions; however, \CIV\ and \SiIV\ column densities may follow single exponential distributions with possibly larger scale heights than previous studies (\citetalias{Savage2009}). The highly ionized gas column density distribution significantly differs from the neutral and low ionization gas below the Perseus arm.

\item Our data do not show strong evidence in the kinematics of the Perseus absorption for rising or falling motions one might expect from Galactic fountain circulation. However, the blending of multiple components within Perseus velocities and the smearing effects of the spectrographs limits our ability to identify such motions. We cannot conclusively confirm or rule out the presence of outflows, inflows, or vertical lag that one might expect in a thickened distribution of gas above a spiral arm.

\item We find the disk-halo interface below the Perseus arm contains a significant amount of WNM gas with associated $\log N(\HI) = 19.5$ compared to the $\log N(\HI)=19.3$ in the WIM. The \SII\ column density in the disk-halo interface 0.5--3.3 kpc below Perseus arm is on average $60$ times higher than \SiIV. However, the column density of total gas associated with the highly ionized (transition temperature) gas is unconstrained in this work.

\item We notice changing ionization conditions from the thin disk to the disk-halo interface gas below the Perseus arm as indicated by the 1) increasing \siIVsII\ values from the thin disk to the disk-halo interface; 2) kinematic alignment of \SII\ \Nav\ profiles with the high ion profiles in the disk-halo interface, which is not the case in the thin disk; and 3) the decreasing scatter in the \cIVsiIV\ values from the thin disk to the disk-halo interface.

\item We observe consistent \siIVsII\ and \cIVsiIV\ values through out the disk-halo interface sightlines. The similarity of the \SII\ and high ion \Nav\ profiles suggest that the high ions in the disk-halo interface below the Perseus arm may reside in turbulent mixing layers at the interfaces between coronal gas at $T>10^6$ K and the \SII\ bearing gas at $T=10^4$ K.

\end{enumerate}


\begin{acknowledgements}

We thank the reviewer for their insightful comments that have helped improving the manuscript. Support for this research was provided by NASA through grant HST-GO-14602 from the Space Telescope Science Institute, which is operated by the Association of Universities for Research in Astronomy, Incorporated, under NASA contract NAS5-26555. This research has made extensive use of the NASA Astrophysics Data System (ADS) Abstract Service and the Centre de Données de Strasbourg (CDS).

\software{This research has made use of adstex (\url{https://github.com/yymao/adstex}), Astropy (\citealt{Astropy2013, Astropy2018, Astropy2022}), Emcee \citep{emcee}, Matplotlib \citep{Hunter2007}, Scipy \citep{scipy}}.

\end{acknowledgements}

\bibliographystyle{aasjournal}
\bibliography{Perseus.bib}

\appendix
\section{Summary Figure}
\label{sec:app-A}

\begin{figure}[h]
\plotone{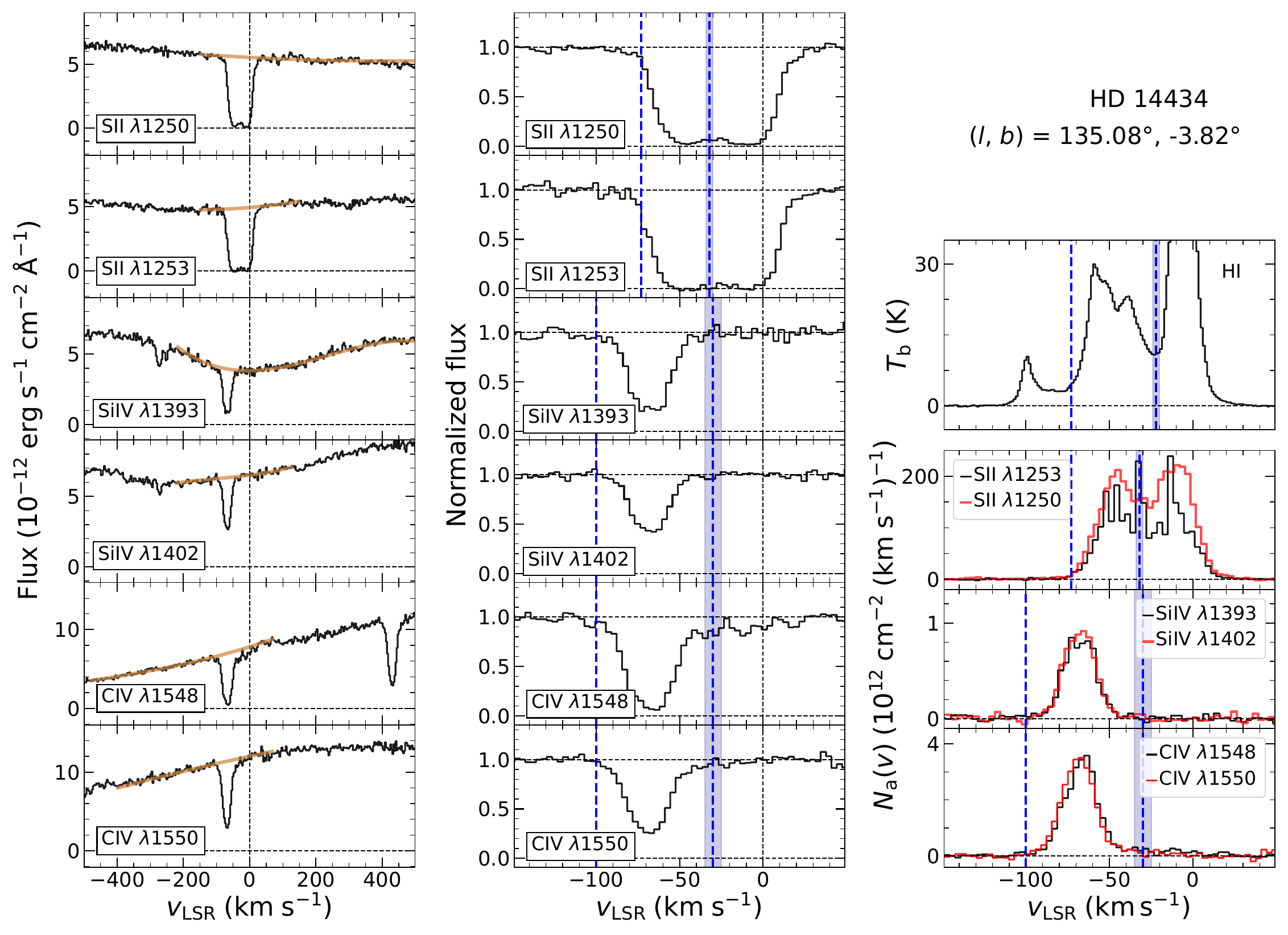}
\caption{Same as Figure \ref{fig:BD+6073}, {\it left}: raw flux with continuum models; {\it middle}: normalized flux, and {\it right}: \HI\ 21 cm brightness temperature and \Nav\ profiles of \SII, \SiIV, and \CIV\ transitions toward each sightline. For details, see Appendix \ref{sec:app-A}.}
\label{fig:all}
\end{figure}

Figure \ref{fig:all} shows the raw flux and continuum model (yellow solid line) ({\it left}) normalized flux ({\it middle}) of \SII\ ($\lambda\lambda$1250,1253), \SiIV\ ($\lambda\lambda$1393, 1402), and \CIV\  ($\lambda\lambda$1548, 1550) against the LSR velocity toward each sample sightline. The \HI\ 21 cm brightness temperature profile, $T_b(v)$, and the \Nav\ profiles of the \SII, \SiIV, and \CIV\ transitions are shown in the right panel. In the top right corner, we present the sightline's name and Galactic coordinates. In the left and middle panels, we mark the absorption from the local gas centered at $v=0$ \kms\ by the black-dash line. In the middle and right panels, the blue-dash lines indicate the adopted Perseus velocity range for each ion (see Table \ref{tab:vel-range}), with the blue shaded region indicating the systematic uncertainties on the velocity upper bound, $v_2$.

\newpage
\section{Co-spatiality of \SII\ and \CIV}
\label{sec:app-B}
\begin{figure}[h]
\plotone{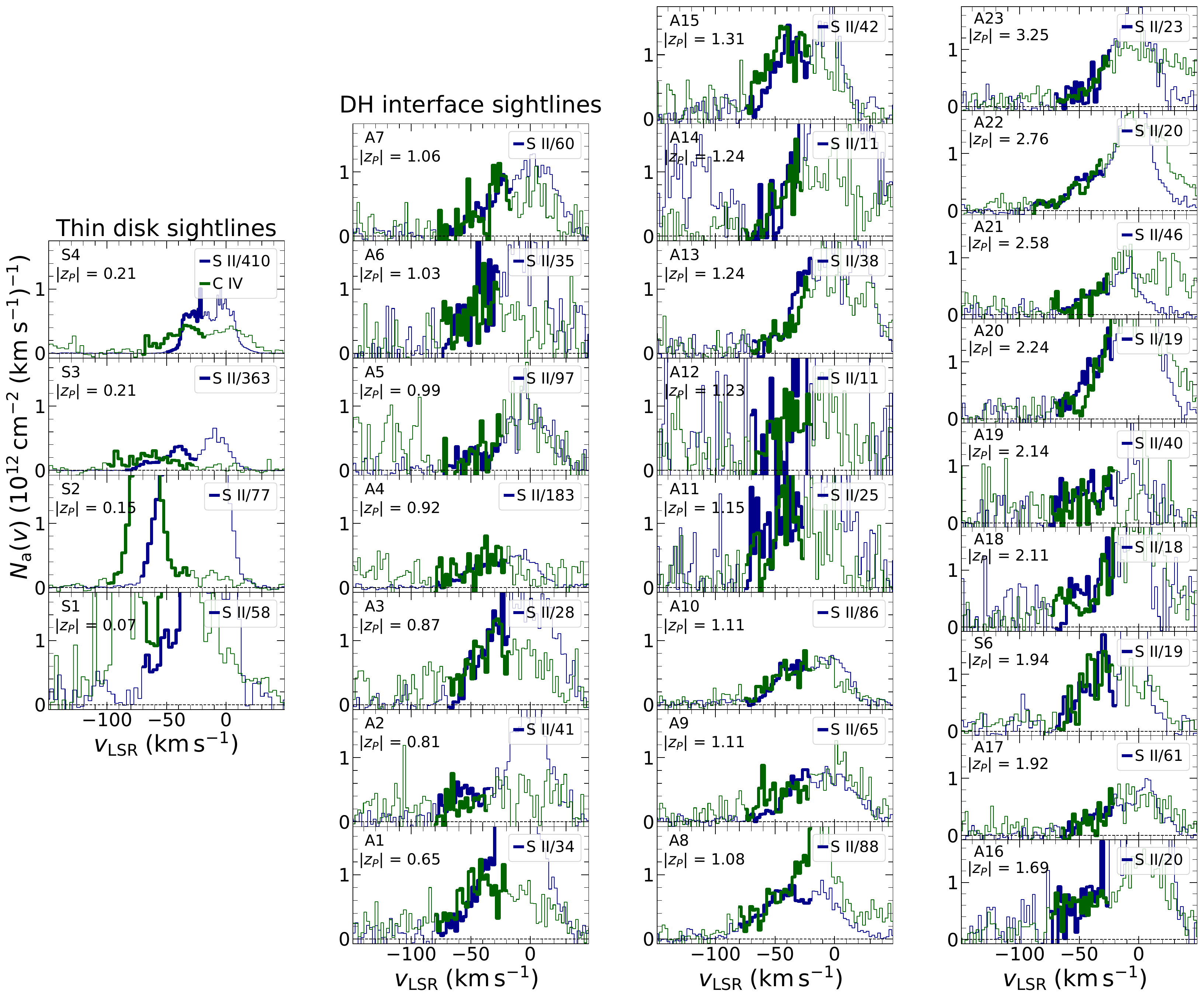}
\caption{Same as Figure \ref{fig:scale-SII-SiIV} and \ref{fig:scale-CIV-SiIV}, toward each sightline, we compare the \CIV\ (green) \Nav\ profile to the \SII\ (blue) profile scaled down by the average \CIV/\SII\ value estimated over the Perseus velocity range for that sightline.}
\label{fig:CIV_SII_full}
\end{figure}

In Figure \ref{fig:CIV_SII_full}, for each sightline, we compare the \CIV\ \Nav\ profile to the \SII\ profile scaled down by the average \CIV/\SII\ value estimated over the Perseus velocity range for that sightline. As in Figure \ref{fig:scale-SII-SiIV}, we find that the scaled \SII\ and \CIV\ \Nav\ profiles kinematically follow each other very well only in the disk-halo interface but very different in the thin disk. This reinstates a physical association between the highly ionized and \SII\ bearing gas in the disk-halo interface below the Perseus arm (see Section \ref{subsec:Nav-comparison} for more details).

\section{Projected Perseus column density along the arm-width as a function of \z}
\label{sec:app-C}

\begin{figure}[h]
    \gridline{\fig{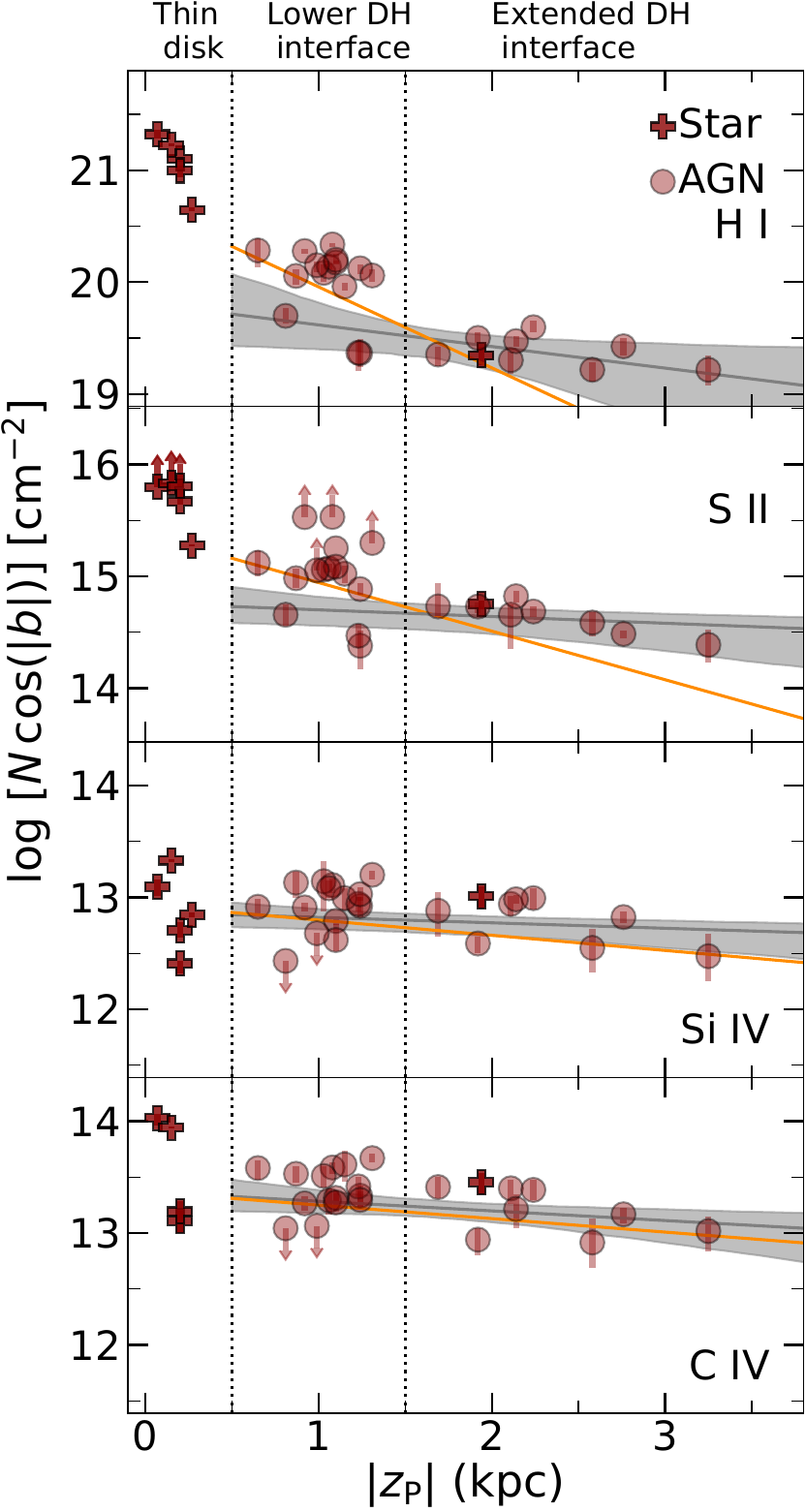}{0.45\columnwidth}{}
              \fig{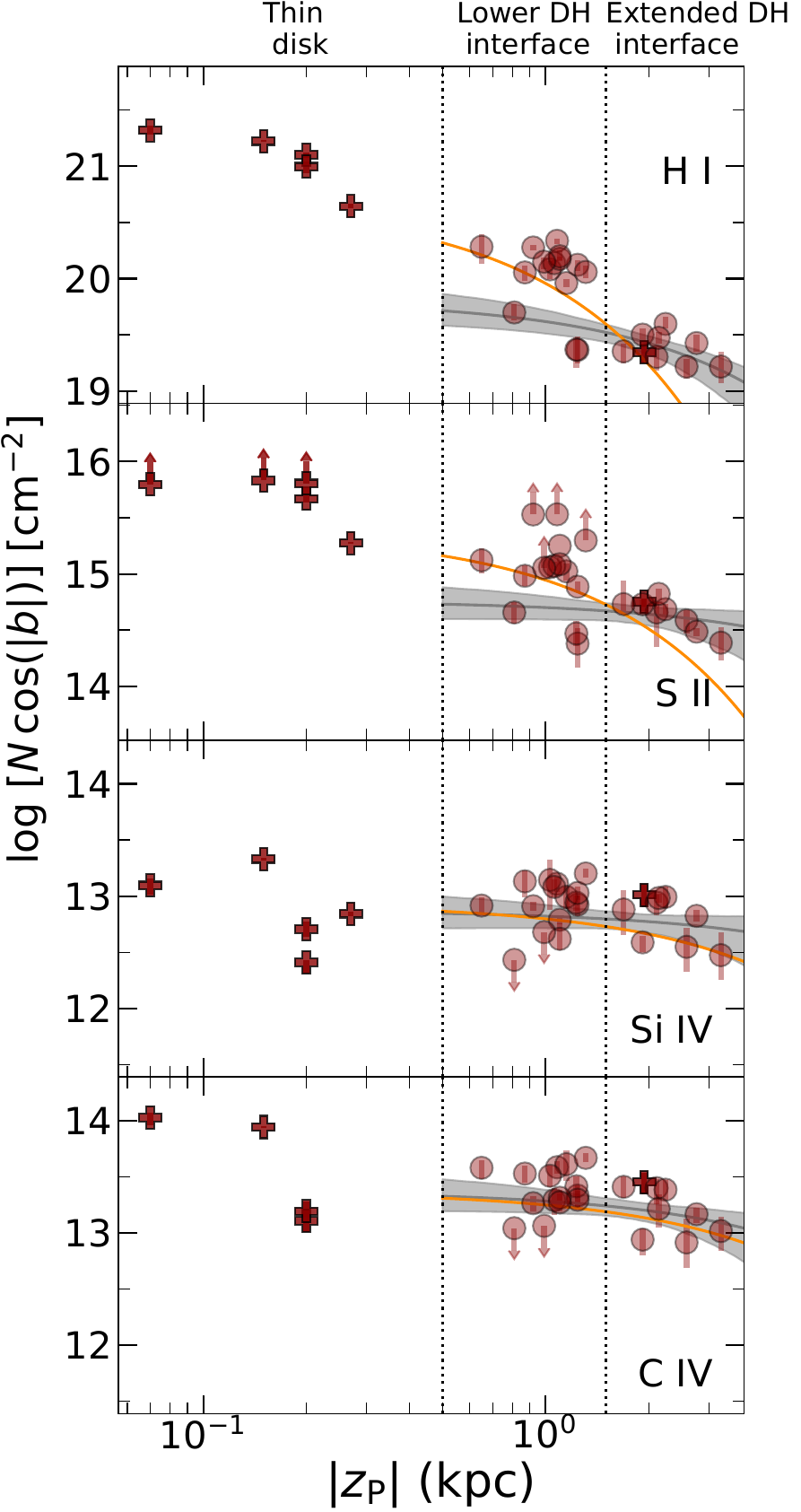}{0.45\columnwidth}{}}
  \caption{Logarithm of \HI, \SII, \SiIV, and \CIV\ projected column densities, $\log [N \cos{|b|}]$, as a function of \z\ ({\it left}) and $\log$ \z\ ({\it right}). The symbols are same as in Figure \ref{fig:colden-zheight}. The gray solid line in each panel indicates the MCMC best fit to the data at \z$>0.5$ kpc assuming a single exponential model, and the gray shaded region indicates the 95\% CI of the MCMC fits. The orange solid line indicates the expected column density distribution with exponential scale heights $h_z$ from literature ($h_z = 0.5$ kpc for \HI\ (\citetalias{Dickey1990}); $1.0$ kpc for \halpha\ emission probing similar gas phase as \SII\ (\citealt{Haffner1999}); 3.2 kpc for \CIV, and 3.4 kpc for \SiIV, (\citetalias{Savage2009})).
  \label{fig:colden-zheight-cosb_corr}}
\end{figure}

In Figure \ref{fig:colden-zheight-cosb_corr}, we show the projected Perseus column density along the arm-width ($\log [N \cos{|b|}]$) vs. mid-arm height \z. The $\log [N \cos{|b|}]$ values are very similar to $\log N$, implying that the effect of latitude dependent variation in the length of sightlines above the Perseus arm is minimal. Consequently, the $\log [N \cos{|b|}]$ vs. \z\ trend in \HI\ and all the ions are consistent with the corresponding $N(z_{\rm P}$) vs.\z\ trend presented in Section \ref{subsec:column_dist}. For example,  \HI\ and \SII\ $\log [N \cos{|b|}]$ sharply drops up to \z\ $=1.5$ kpc, however, the decline is relatively shallow at \z\ $>1.5$ kpc and $\log [N \cos{|b|}]$ values significantly deviate from the flat-slab model predictions (the orange lines). In contrast, the high ion $\log [N \cos{|b|}]$ decline is very shallow over the whole range of \z, however, their $\log [N \cos{|b|}]$ vs. \z\ distributions might be consistent with 3--4 kpc scale heights of \citetalias{Savage2009}.

\section{Comparison of the Column Densities Toward the Halo Stars and AGNs}

Similar to Figure \ref{fig:total_col_south}, we present a comparison of the \CIV\ column densities toward halo stars from \citet{Bish2021}, our halo star, and the AGNs from \cite{Qu2022}. In addition to the column densities, we show linear regression fits for the halo stars (south: grey dashed line, north: teal dashed line) and AGNs (south: teal solid line, north: teal solid line). The top right panel displays the residuals (observed $N \sin{|b|}$ $-$ the linear model) as a function of stellar distance $D_{\rm{star}}$ for halo stars in both hemispheres. The middle and bottom right panels show the residuals as a function of $|b|$ for AGNs and halo stars in southern and northern hemisphere, respectively.
As discussed in Section \ref{subsec:total-column}, the residual distributions of the halo stars are statistically consistent with the AGN samples. Additionally, the halo star residuals exhibit no dependence on either $|b|$ or $D_{\rm{star}}$. These observations provide compelling statistical evidence that the $N \sin{|b|}$ versus $|b|$ relationship toward halo stars is consistent with that observed toward AGNs.

\label{sec:app-D}

\begin{figure}[h]
\plotone{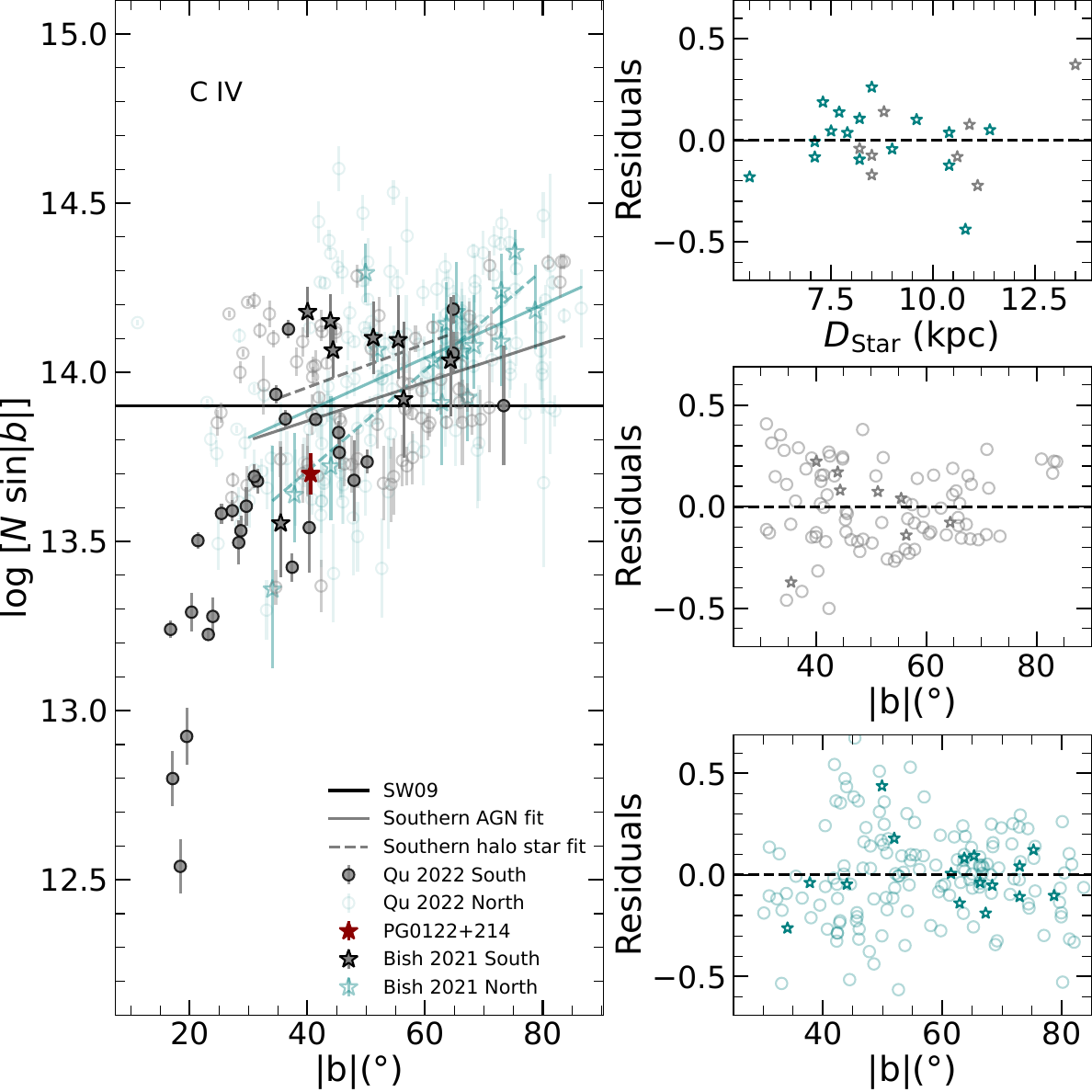}
\caption{ Similar to Figure \ref{fig:total_col_south}, the left panel displays $N \sin{|b|}$ as a function of $|b|$ in the halo star \citep{Bish2021} and AGN sightlines \citep{Qu2022} along with the linear regression fits. The top right panel displays the residuals (observed $N \sin{|b|}$ $-$ the linear model) as a function of stellar distance $D_{\rm{star}}$ for halo stars in both hemispheres. The middle and bottom right panels show the residuals as a function of $|b|$ for AGNs and halo stars in southern and northern hemisphere, respectively.
\label{fig:total_col_south_updated}}
\end{figure}

\end{document}